\documentclass[11pt, oneside,reqno]{amsart}   	
\usepackage[margin=1.5cm]{geometry}                		
\usepackage{amsaddr}

\usepackage{graphicx}				
\usepackage{tikz}
\usetikzlibrary{decorations.pathmorphing}
\tikzset{snake it/.style={decorate, decoration=snake}}
\usepackage{cite}
\usepackage[T1]{fontenc}

\usepackage{framed}
\definecolor{shadecolor}{gray}{0.95}
\usepackage{hyperref}
\usepackage{xfrac}
\usepackage{amsmath,amssymb,amsthm}

\renewcommand{\Re}{\operatorname{Re}}
\renewcommand{\Im}{\operatorname{Im}}
 \newcommand{\rmi}{\mathrm{i}}
 \newcommand{\rmd}{\mathrm{d}}
 \newcommand{\eps}{\varepsilon}
 \newcommand{\rme}{\mathrm{e}}
 \newcommand{\tr}{\mathrm{tr}}
  \hypersetup{
    colorlinks=true,
    linkcolor=black,
    citecolor=black,
    filecolor=black,
    urlcolor=blue,
}

\newtheorem{result}{Result}

\begin{document}

\title[Symmetry-resolved entanglement entropy in critical chains]{Symmetry-resolved entanglement entropy in critical free-fermion chains.}
\author{Nick G. Jones}
\address{Mathematical Institute, University of Oxford, Oxford, OX2 6GG, UK\\
and the Heilbronn Institute for Mathematical Research, Bristol, UK}
\email{nick.jones@maths.ox.ac.uk}
\thanks{The published version of this article is J. Stat. Phys., 188, 28 (2022); \href{https://doi.org/10.1007/s10955-022-02941-3}{https://doi.org/10.1007/s10955-022-02941-3}.}
\begin{abstract}
The symmetry-resolved R\'enyi entanglement entropy is the R\'enyi entanglement entropy of each symmetry sector of a density matrix $\rho$. This experimentally relevant quantity is known to have rich theoretical connections to conformal field theory (CFT). For a family of critical free-fermion chains, we present a rigorous lattice-based derivation of its scaling properties using the theory of Toeplitz determinants. 
We consider a class of critical quantum chains with a microscopic U(1) symmetry; each chain has a low energy description given by $N$ massless Dirac fermions. For the density matrix, $\rho_A$, of subsystems of $L$ neighbouring sites we calculate the leading terms in the large $L$ asymptotic expansion of the symmetry-resolved R\'enyi entanglement entropies. This follows from a large $L$ expansion of the charged moments of $\rho_A$; we derive $\tr(\rme^{\rmi \alpha Q_A} \rho_A^n)~=~a \rme^{\rmi \alpha \langle Q_A\rangle}  (\sigma L)^{-x}(1+O(L^{-\mu}))$, where $a, x$ and $\mu$ are universal and $\sigma$ depends only on the $N$ Fermi momenta. We show that the exponent $x$ corresponds to the expectation from CFT analysis. The error term $O(L^{-\mu})$ is consistent with but weaker than the field theory prediction $O(L^{-2\mu})$. However, using further results and conjectures for the relevant Toeplitz determinant, we find excellent agreement with the expansion over CFT operators. 
\end{abstract}
\maketitle

\section{Introduction}
Entanglement is a fundamental concept in quantum information theory and quantum many body physics \cite{Amico08,Laflorencie16,Zeng2019}. One important context is entanglement in ground states of many body systems. A key insight is the area law for entanglement for gapped phases of matter; of importance for tensor network descriptions of these states \cite{Cirac21}.
A major result was the proof of the area law in general one-dimensional systems \cite{Hastings07,Arad13}. This area law is violated at phase transitions; entanglement properties are then closely related to the nature of the quantum critical point. For example, in one-dimensional spin chains with low-energy conformal field theory (CFT) description, the von Neumann entanglement entropy, $S(L)$, of a subsystem consisting of $L$ neighbouring spins (away from any boundary) obeys a universal scaling law \begin{align}S(L) = \frac{c}{3} \log(L) +O(1), \label{eq:entanglementscaling}\end{align} where $c$ is the central charge of the CFT \cite{Vidal03,Calabrese04,Korepin2004,Calabrese09}. 

More recently, \emph{symmetry-resolved} analogues of the usual entanglement entropies have been introduced and analysed \cite{Song12,Laflorencie14,Goldstein18,Xavier18}. Roughly speaking, one block diagonalises the reduced density matrix of the subsystem into different sectors of fixed charge and then finds the entropy of each sector. This is a natural problem, and has experimental relevance \cite{Lukin19,Azses20,Vitale2021,Neven21}.
One theoretical result is the \emph{equipartition of entanglement}: for CFTs with U(1) symmetry, the leading order behaviour is independent of the symmetry sector \cite{Xavier18}. The leading charge-dependent contributions to the entanglement were identified in exact calculations in free-fermion chains in Ref.~\cite{Bonsignori19}. There is a growing literature on this topic; while a full review is beyond our scope, we note works based on CFT and the replica trick \cite{Goldstein18,Xavier18,Murciano2021}, approaches combining CFT and lattice form factors \cite{Estienne21} and lattice results based on Toeplitz determinant theory \cite{Bonsignori19,Fraenkel20,Ares22}. There are moreover extensions to more general quantum field theories \cite{Horvath2020,Murciano2020,Horvath22}, spacetimes with dimension greater than two \cite{Tan2020,Murciano2020b}, symmetry-resolved negativity \cite{Cornfeld2018,Murciano2021c}, symmetry-resolved relative entropies \cite{Capizzi2021,Chen2021} as well as the out-of-equilibrium behaviour of the symmetry-resolved entanglement following a quench \cite{Fagotti08,Feldman19,Parez21,Parez21b,Fraenkel21,Parez22}. Note that in the literature, the concept of symmetry-resolved entanglement (in particular, for the setting of a U(1) symmetry generated by the particle number) is closely related to the concepts of configurational entanglement, entanglement of particles and operational entanglement \cite{Lukin19,Wiseman03,Melko16}.

In this paper we analyse symmetry-resolved entanglement entropies in a class of critical quantum chains with U(1) symmetry. The low energy description of these chains is a theory of $N$ massless Dirac fermions, each with associated momentum $k_j$ and velocity $v_j$ for $1\leq j \leq N$. For $N=1$ this is a CFT with $c=1$ (the usual compactified boson at the free-fermion radius). For an example in this class (a nearest neighbour tight-binding model, Jordan-Wigner dual to the spin-1/2 XX model \cite{Sachdev01}) entanglement entropies \cite{Jin2004} and symmetry-resolved entanglement entropies \cite{Bonsignori19,Fraenkel20,Estienne21} have been analysed previously using lattice methods. For $N>1$, we have a low-energy CFT description only when each of the $v_j$ are equal: an $\mathrm{SO}(2N)_1$ Wess-Zumino-Witten (WZW) model with $c=N$ \cite{Lahtinen15}. For general velocities, the case $N>1$ can still be understood using CFT concepts: the scaling law \eqref{eq:entanglementscaling} holds with $c=N$  \cite{Keating2004} and one can identify low-energy descriptions of lattice operators by taking sums and products of CFT operators from each of the different sectors \cite{Shankar94,Bogoliubov87,Izergin89,Hutchinson2016,Jones2019} (each sector being a copy of the $c=1$ CFT). Moreover, we give an explicit argument that for a family of $N=2$ Hamiltonians that are not described at low energies by a CFT (since they have different Fermi velocities), the ground state has a parent Hamiltonian that is described at low energies by a CFT. Hence, ground state properties such as entanglement entropies should agree exactly with CFT results. We expect this to hold more generally within the family of models considered. Indeed, this point was made in Ref.~\cite{Izergin89} for general Bethe Ansatz solvable models.

Consider now a subsystem of the critical chain consisting of $L$ neighbouring sites. Our main result consists of the leading terms in the large $L$ asymptotic expansion of the symmetry-resolved R\'enyi entropies. In order to derive this, we calculate the leading term, and give an estimate for the error, in the asymptotic expansion of the charged moment (defined below); taking a Fourier transform then leads to the symmetry-resolved R\'enyi entropies. Fundamental to our proof is the asymptotic analysis of Toeplitz determinants with Fisher-Hartwig singularities \cite{Deift2013}. These methods were applied to the computation of entanglement entropies in the XX model by Jin and Korepin \cite{Jin2004} and then to the class of chains we consider here by Keating and Mezzadri \cite{Keating2004}. See also \cite{Kadar10,Ares15} for further examples and references. This method was generalised by Bonsignori, Ruggiero and Calabrese \cite{Bonsignori19} and by Fraenkel and Goldstein \cite{Fraenkel20} to the computation of symmetry-resolved entropies in the XX model. We largely follow their method to prove our results for the broader class of models considered here. However, an important point is that the analysis given in those papers relies on a conjectured expansion of the relevant Toeplitz determinant to identify subleading terms (in fact the dominant subleading terms were derived by Kozlowski \cite{Kozlowski08}, but not in a form that we can use in our proof). We set up the problem differently so that we can use a rigorous and sufficiently uniform bound from Ref.~\cite{Deift2014}. This approach does, however, lead to a weaker bound on the correction term than expected---we address this in our discussion. While our results for $N=1$ largely reproduce the formulae already in the literature, we consider our analysis to be of interest also in that case since we can give a rigorous estimate of the errors at each stage. 

Motivated by our analysis of the symmetry-resolved R\'enyi entropies, and the work of \cite{Calabrese10,Bonsignori19,Fraenkel20}, we give an extended discussion of the calculation of subleading terms in the charged moments using both CFT and Toeplitz determinant methods. While not leading to a rigorous result, we find that applying these two different techniques gives the same estimate for the algebraic decay of the subleading term (which is generally the same as the $N=1$ XX model case \cite{Bonsignori19,Fraenkel20}), and also gives the corresponding oscillatory terms (going beyond the XX model case) that allow us to match CFT operators to \emph{Fisher-Hartwig representations} of certain functions. We give an exact expression for the subleading term in all cases, based on certain assumptions about the asymptotic expansion.  

The outline of the paper is as follows. In Section \ref{sec:results} we define the model and quantities of interest before stating our results. This is followed in Section \ref{sec:proof} by the corresponding proofs. In Section \ref{sec:discussion} we discuss our results, make connections to corresponding field theoretical calculations and analyse subleading correction terms. Note that Sections \ref{sec:CFT}--\ref{sec:decoupled} in the discussion can be read independently of the proofs. Section \ref{sec:FH} analyses the subleading terms from Toeplitz determinant theory, and relies on certain definitions given in Section \ref{sec:asymptotics}. Finally we discuss further avenues of research.

\section{Statement of results}\label{sec:results}
\subsection{A class of models}\label{sec:class}
Consider a chain of $M$ sites, each with a spinless fermionic degree of freedom $c_n$, such that $\{c_n,c_m\} =0$ and  $\{c^\dagger_n,c^{\vphantom \dagger}_m\} =\delta_{nm}$. 
We will take the limit $M\rightarrow\infty$ and examine the class of fermionic hopping Hamiltonians:
\begin{align}
H = -\sum_{n \in \mathrm{sites}} \Bigg(t_0  c^\dagger_n c^{\vphantom{\dagger}}_{n} + \sum_{\alpha=1}^R t_\alpha(c^\dagger_n c^{\vphantom{\dagger}}_{n+\alpha} +c^\dagger_n c^{\vphantom{\dagger}}_{n-\alpha}) \Bigg) \qquad t_\alpha \in \mathbb{R}.
\end{align}
This class has a U(1) symmetry generated by $Q=\sum_{n \in \mathrm{sites}}  c^\dagger_n c^{\vphantom{\dagger}}_{n} $. We can diagonalise by Fourier transformation:
\begin{align}
H = -\sum_{k} \underbrace{\left(  t_0 + 2\sum_{\alpha=1}^R t_\alpha \cos(k\alpha)\right)}_{f(\rme^{\rmi k})} c^\dagger_k c^{\vphantom{\dagger}}_{k}. \label{eq:model}
\end{align}
The ground state has filled modes with momentum $k$ such that $f(\rme^{\rmi k})>0$. If the function $f(z)$ has no zeros on the unit circle, then the ground state is trivial. Zeros on the unit circle generically occur\footnote{Real zeros at $k=0$ or $k=\pi$ and with multiplicity one are not consistent with the condition $f(z)=f(1/z)$ that is satisfied by the class of models above. The free-fermion model (outside of the class above) corresponding to $f(\rme^{\rmi k}) \propto \sin(k)$ could be studied using the techniques here. The general analysis is in fact unchanged if all zeros of $f(z)$ on the unit circle have odd multiplicity (simply use the result where the corresponding zeros have multiplicity one). Models with some zero on the unit circle having even multiplicity correspond to multicritical points (the multicritical point in the phase diagram of the XY model is an example). These zeros are `removable'---for further discussion see Refs. \cite{Hutchinson2016,Jones2019,Jones21}.} in conjugate pairs and with multiplicity one. If we have $N$ such pairs then we can linearise the dispersion about each point to find the low energy behaviour described in the introduction \cite{Shankar94}. We denote\footnote{In the case $N=1$ we will also use the notation $k_F$ for the Fermi momentum.} the independent momenta by $0<k_1<\dots<k_N <\pi$. Then:
\begin{align}
f(\rme^{\rmi k})=\pm h(k) \prod_{j=1}^N \sin\left(\frac{k-k_j}{2}\right)\sin\left(\frac{k+k_j}{2}\right),  \label{eq:standardform} \end{align}
where $ h(k)=h(-k) $ is real and strictly positive---the ground state is independent of $h(k)$, and while the excitation energies (and Fermi velocities $v_j= \lvert\partial_k f(\rme^{\rmi k_j})\rvert$) depend on $h(k)$, it will play no role in our analysis. Indeed, the symmetry-resolved entropies are properties of the ground state and are otherwise independent of the Hamiltonian. 

Using the Jordan-Wigner transformation this model is equivalent to the spin-1/2 chain:
\begin{align}
H_{\mathrm{spin}} = \frac{1}{2} \sum_{n \in \mathrm{sites}} \Bigg(t_0 Z_n  - \sum_{\alpha=1}^R t_\alpha\Big(X_n \left(\prod_{m=n+1}^{n+\alpha-1} Z_m\right) X_{n+\alpha}+Y_n \left(\prod_{m=n+1}^{n+\alpha-1} Z_m\right) Y_{n+\alpha}\Big) \Bigg) \qquad t_\alpha \in \mathbb{R},
\end{align}
where $X_n, Y_n$ and $Z_n$ are the usual Pauli operators on site $n$. This is a family of generalised cluster models \cite{Suzuki71,Keating2004,Verresen17}, with U(1) symmetry generated by $Q=\sum_{n \in \mathrm{sites}}  Z_n$. Such chains appear as transitions between certain symmetry-breaking and symmetry-protected topological (SPT) phases \cite{Verresen17,Verresen2018}. We note that the symmetry-resolved entanglement of certain ground states in such gapped SPT phases was studied in Ref.~\cite{Azses20b}.
 
\subsection{Symmetry-resolved entropies, charged moments and symmetry-resolved moments}
Consider a bipartition of our critical chain into two subsystems $A$ and $B$, where $A$ consists of $L$ neighbouring sites. The density matrix of the subsystem is given by $\rho_A=\tr_B (\rho) $ where $\rho$ is the ground state. The U(1) charge acts locally as $Q=Q_A+Q_B$; let $\Pi_q$ be the projector onto the eigenspace of $Q_A$ with eigenvalue $q$. Then $\rho_A$ can be decomposed as $\rho_A = \oplus_q p(q) \rho_A(q)$ where $p(q) = \tr(\Pi_q \rho_A)$ and for $p(q) \neq0$ we define normalised density matrices $\rho_A(q) = {(\Pi_q \rho_A\Pi_q)}/p(q)$. Then we can define the symmetry-resolved R\'enyi entropies by:
\begin{align}
S_n(q) = \frac{1}{1-n} \log \tr\Big(\rho_A (q)^n\Big)
\end{align}
and the symmetry-resolved von Neumann entropy by $S(q) = -\tr\Big(\rho_A(q) \log (\rho_A(q))\Big)$. Note that $S(q)$ is the limit of $S_n(q)$ as $n\rightarrow 1$.

In order to calculate the entropy, we first define the symmetry-resolved moment:
\begin{align}
\mathcal{Z}_n(q) = \tr\Big(\Pi_q \rho_A^n\Big),
\end{align}
then $S_n(q) = \Big(\log\big(\mathcal{Z}_n(q)\big) - n\log\big(\mathcal{Z}_1(q)\big) \Big)/(1-n)$.
The symmetry-resolved moment  is the Fourier transform of the charged moment:
\begin{align}
\mathcal{Z}_n(q) &= \frac{1}{2\pi}  \int_{-\pi}^\pi \rme^{-\rmi q \alpha} Z_n(\alpha) \rmd \alpha\\
\textrm {where} \quad Z_n(\alpha) &=  \tr\Big(\rme^{\rmi \alpha Q_A} \rho_A^n\Big).
\end{align}
For definiteness $Q_A = \sum_{j=1}^L c_j^\dagger c^{\vphantom\dagger}_j$ is the number of particles in the region $A$. This is the key step to allow the calculation of the symmetry-resolved entropies, since we can calculate $Z_n(\alpha)$ as a contour integral where we have asymptotics for the integrand \cite{Goldstein18,Xavier18,Bonsignori19,Fraenkel20}. Note that for $n=1$ the charged moment is equal to the full-counting-statistics generating function that has been studied before, at least for the XX model \cite{Abanov11,Ivanov13}. More generally, this quantity appears (in the obvious way) in defining charged R\'enyi entropies in a holographic setting \cite{Belin13,Estienne21}.
\subsection{Results}
To state our results, define the quantity $L_\sigma = \sigma L$, where, given the momenta $k_1,\dots,k_N$, the (inverse) length scale\footnote{More carefully: the length scale is $a/\sigma$ where $a$ is the lattice spacing.} $\sigma$ is defined as:
 \begin{align}
 \sigma = \Bigg(\prod_{r=1}^N 2 \sin(k_r) \prod_{1\leq r<s\leq N}  
\Bigg(\frac{\sin^2 (\frac{k_r+k_s}{2})}{\sin^2 (\frac{k_r-k_s}{2})} \Bigg)^{(-1)^{r+s}}\Bigg)^{1/N}.\label{eq:sigma}
\end{align}
We also define the function \cite{Bonsignori19,Fraenkel20}:
\begin{align}
 \Upsilon(n,\alpha)&= \rmi n \int_{-\infty}^\infty \left(\tanh(\pi w) - \tanh(n \pi w +\rmi \alpha/2) \right)\log \frac{\Gamma(1/2+\rmi w)}{\Gamma(1/2-\rmi w)} \rmd w \label{def:upsilon}.
 \end{align}
 For $n\in\mathbb{Z}_+$,  $\Upsilon(n,\alpha)$ has the closed form \cite{Estienne21}:
\begin{align}
\Upsilon(n,\alpha) &=2\sum_{m=0}^{n-1} \log\Bigg( G\Big(1-\frac{\alpha}{2\pi n}+\frac{2m+1-n}{2n} \Big) G\Big(1+\frac{\alpha}{2\pi n}+ \frac{2m+1-n}{2n}  \Big)\Bigg),\label{eq:upsilon}\end{align}
where $G(z)$ is the Barnes G-function  \cite{NIST:DLMF}. Discussion and alternative expressions for $ \Upsilon(n,\alpha)$ are found in Section \ref{sec:upsilon}.

From the characterisation of the ground state of \eqref{eq:model} in momentum space, the extensive mean charge in the subsystem $A$ is given by:
\begin{align}
\langle Q_A\rangle = \begin{cases} \frac{\sum_{r=1}^N (-1)^{N-r} k_r}{\pi} L \qquad  &f(\rme^{\rmi \pi})<0\\ 
\frac{\pi - \sum_{r=1}^N (-1)^{N-r} k_r}{\pi} L  \qquad&f(\rme^{\rmi \pi})>0.
\end{cases} 
\end{align}
We now state our results. We emphasise that related formulae for the XX model case are found in \cite{Bonsignori19,Fraenkel20,Estienne21}.

\begin{shaded}
\begin{result}[Charged moments]\label{resultone}
Consider a system in our class of the form \eqref{eq:standardform}, with a low energy theory of $N$ complex fermions, and take a subsystem consisting of $L$ neighbouring sites. The charged moment of the subsystem has the following asymptotic expansion as $L \rightarrow \infty$:
\begin{align}
&Z_n(\alpha,L) = L_\sigma^{-\frac{N}{6}\left(n-\frac{1}{n}\right)-\frac{2N}{n} \left(\frac{\alpha}{2\pi}\right)^2} \rme^{\rmi \alpha \langle Q_A\rangle}\rme^{N \Upsilon(n,\alpha)}  \Big(1+O\left(L^{-\mu(n,\alpha)}\right)\Big) \qquad -\pi<\alpha<\pi,
\end{align}
where $0<n\in\mathbb{R}$ and
\begin{align} 
\mu(n,\alpha)= \min\Bigg\{\frac{1}{2},  \frac{1}{n} \left(1-\frac{\lvert\alpha\rvert}{\pi}\right) \Bigg\}.\label{eq:mu}\end{align}
\end{result}
\end{shaded}
Note that $\mu(n,\alpha)$ tends to zero as $\alpha \rightarrow \pm\pi$; in this limit the correction term is not necessarily subdominant. This is as expected from both the CFT analysis and the analysis using Toeplitz determinant theory. Note also that for $n\geq 2$, we simply have  $\mu(n,\alpha) =\frac{1}{n} \left(1-\frac{\lvert\alpha\rvert}{\pi}\right)$. For $n=1$ and $\alpha =0$, we can improve the bound on the error term, see discussion in Appendix \ref{app:errorterm}.

To state our next result, we expand $ \Upsilon(n,\alpha)$ as follows:
\begin{align}
 \Upsilon(n,\alpha)=  \Upsilon(n) +\alpha^2 \gamma_2(n) + \alpha^4 \gamma_4(n) + \epsilon(n,\alpha) \label{def:upsilonexpansion}
\end{align}
where $\epsilon(n,\alpha) = O(\alpha^6)$ and 
\begin{align}
\gamma_2(n)&=\frac{\rmi n}{4} \int_{-\infty}^\infty \left(\tanh ^3(\pi  n w)-\tanh (\pi  n w)\right)\log \frac{\Gamma(1/2+\rmi w)}{\Gamma(1/2-\rmi w)} \rmd w \nonumber\\
\gamma_4(n)&=\frac{\rmi n}{192} \int_{-\infty}^\infty \frac{\sinh(3 \pi  n w)-11\sinh (\pi  n w)}{\cosh^5( \pi n w)}\log \frac{\Gamma(1/2+\rmi w)}{\Gamma(1/2-\rmi w)} \rmd w.\label{eq:gamma2,4}
\end{align} 
The above integral formulae hold for any $0<n\in \mathbb{R}$. For $n\in \mathbb{Z}_+$ we have:
\begin{align}
\gamma_2(n) = 2\sum_{m=0}^{n-1} \frac{1}{4n^2\pi^2} \left(-1+\psi\Big(1+\frac{2m+1-n}{2n}\Big)+\frac{2m+1-n}{2n}\psi^{(1)}\Big(1+\frac{2m+1-n}{2n}\Big) \right)\nonumber\\
\gamma_4(n) = 2\sum_{m=0}^{n-1} \frac{1}{192n^4\pi^4} \left(3\psi^{(2)}\Big(1+\frac{2m+1-n}{2n}\Big)+\frac{2m+1-n}{2n}\psi^{(3)}\Big(1+\frac{2m+1-n}{2n}\Big) \right) \label{eq:gamma2,4n}
\end{align}
where $\psi(x)=\Gamma'(x)/\Gamma(x)$ is the polygamma function, with derivatives $\psi^{(n)}(x)$.
\begin{samepage}
\begin{shaded}
 \begin{result}[Symmetry-resolved moments]\label{resulttwo}
 Consider a system in our class of the form \eqref{eq:standardform}, with a low energy theory of $N$ complex fermions, and take a subsystem consisting of $L$ neighbouring sites. 
 Define the quantity 
\begin{align}
a(n,L)&=\frac{N}{2n\pi^2}\left( \log(  L_\sigma) -2n\pi^2\gamma_2(n) \right)= O(\log L) .\end{align}
Then, in the sector with charge $q= q_\Delta+ \langle Q_A\rangle$, where $q_\Delta = O(1)$, the symmetry-resolved 
 moment of the subsystem has the following asymptotic expansion as $L \rightarrow \infty$:
  \begin{align}
\mathcal{Z}_n( q) &= \frac{Z_n(0,L)}{\sqrt{4\pi a(n,L)}} \exp\left(- \frac{ q_\Delta^2}{4a(n,L)}\right) \Bigg(1+\frac{3}{4}\frac{N\gamma_4(n)}{a(n,L)^2}+ O(\log{(L)}^{-3})\Bigg).
\end{align}
\end{result}\end{shaded}
\end{samepage}
This second result leads directly to an expansion for the symmetry-resolved entanglement entropies of the subsystem:
 \begin{shaded}
 \begin{result}[Symmetry-resolved entropy]\label{resultthree}
  With the same conditions as Result \ref{resulttwo}, the symmetry-resolved R\'enyi entropy has the following asymptotic expansion as $L\rightarrow\infty$:
  \begin{align}
S_n(q) &= \frac{N}{6}\left(1+\frac{1}{n}\right)\log( L_\sigma)  + \frac{N}{1-n} \Upsilon (n)   -\frac{1}{2}\log\left(\frac{2N}{\pi}\log(L)\right)+ \frac{1}{2(1-n)}\log(n) \nonumber\\&+\log(L)^{-1}\Bigg(\frac{n\pi^2\Big(\gamma_2(n)-\gamma_2(1)\Big)}{(1-n)}-\frac{1}{2}\log{(\sigma)}\Bigg)\nonumber\\&+\log(L)^{-2}\Bigg[\frac{n\pi^2}{(1-n)}\Bigg(\pi^2(n\gamma_2(n)^2-\gamma_2(1)^2)+\log(\sigma) (\gamma_2(1)-\gamma_2(n))+\frac{3\pi^2}{N}(n\gamma_4(n)-\gamma_4(1))\Bigg)\nonumber\\&+\frac{1}{4}\log{(\sigma)}^2  + q_\Delta^2\frac{ n \pi^4}{N(1-n)}(\gamma_2(1)-n\gamma_2(n))\Bigg] 
 +O(\log{(L)}^{-3}).\end{align}\end{result}
 \end{shaded}
 As emphasised in \cite{Bonsignori19}, we see that the equipartition of entanglement breaks at order $\log(L)^{-2}$. We can take the limit $n\rightarrow 1$ to find the asymptotics of the symmetry-resolved von Neumann entanglement entropy,
 \begin{align}
S(q) &= S  -\frac{1}{2}\log\left(\frac{2N}{\pi}\log(L)\right)-\frac{1}{2} -\log(L)^{-1}\Bigg(\pi^2\gamma_2'(1)+\frac{1}{2}\log{(\sigma)}\Bigg)\nonumber
\\&+\log(L)^{-2}\Bigg[-2\pi^4 \gamma_2(1)\gamma_2'(1)+\pi^2\gamma_2'(1)\log(\sigma)-\frac{3\pi^4}{N}(\gamma_4(1)+\gamma_4'(1))+\frac{1}{4}\log{(\sigma)}^2 \nonumber\\& + q_\Delta^2\frac{  \pi^4}{N}(\gamma_2(1)+\gamma_2'(1))\Bigg] 
 +O(\log{(L)}^{-3}).\end{align}
Here $S$ is the usual von Neumann entropy calculated in \cite{Keating2004}, and the derivatives $\gamma_2'(1)$ and $\gamma_4'(1)$ can be given as integrals using \eqref{eq:gamma2,4}.  Numerically, $\gamma_2'(1) \simeq 0.0546$ and $\gamma_4'(1) \simeq 0.00154$ \cite{Bonsignori19,Mathematica}. From \eqref{eq:gamma2,4n}, we have analytic expressions $\gamma_2(1)=-(1+\gamma_E)/2\pi^2$ \cite{Bonsignori19,Xavier18,Estienne21}, where $\gamma_E$ is Euler's constant, and $\gamma_4(1) = \psi^{(2)}(1)/32\pi^4$.

Following Ref.~\cite{Bonsignori19}, note that the $O(1)$ term can be understood as follows. Since the density matrix is $\rho_A = \oplus_q p(q)\rho_A(q)$, we can write $S = \sum_q p(q) S(q) - \sum_q p(q)\log(p(q))$. The first term corresponds to \emph{configurational entanglement}, while the second is the \emph{fluctuation entanglement}.  
Since $ \mathcal{Z}_1(q) = p(q) $, we can see using Result \ref{resulttwo} that the fluctuation entanglement $-\sum_q p(q)\log(p(q))$ is equal to $\frac{1}{2}\Big(1+\log\left(\frac{2N}{\pi}\log(L)\right)\Big)+o(1)$. Hence, at leading order, the symmetry-resolved entanglement entropy obeys equipartition and is equal to the configurational entanglement---this is natural since the symmetry-resolved entanglement entropy corresponds to the entanglement entropy in a fixed charge sector and so there is no fluctuation entanglement.

 \section{Analysis}\label{sec:proof}
 In this section we give proofs of the above results. Section \ref{sec:proofone} leads to Result \ref{resultone}. This is the most involved part of the analysis. We give various results relating to the term $\Upsilon(n,\alpha)$ in Section \ref{sec:upsilon}. In Section \ref{sec:prooftwo}, we reach Result \ref{resulttwo}---this is straightforward analysis of a Gaussian integral, but does require the error term in Result \ref{resultone} to justify the expansion. Finally, the steps to write down the asymptotics for the symmetry-resolved entropies are given in Section \ref{sec:proofthree}. 
\begin{samepage} \subsection{Asymptotics of the charged moments}\label{sec:proofone}
\subsubsection{Set up}
In the ground state of our class of models we have the two-point correlator:
\begin{align}
\langle c^\dagger_m c^{\vphantom\dagger}_n \rangle = \frac{1}{2\pi}\int_0^{2\pi} \rme^{-\rmi k (m-n)} \frac{(\mathrm{sign}[f(\rme^{\rmi k})] +1)}{2} \rmd k.
\end{align}\end{samepage}
Consider the $L\times L$ correlation matrix defined by $C_{mn} = \langle c^\dagger_m c^{\vphantom\dagger}_n \rangle$ for $1\leq m,n\leq L$, and define its $L$ eigenvalues by $\frac{1+\nu_j}{2}$ for $1\leq j \leq L$. The $\nu_j$ are eigenvalues of a Toeplitz matrix, $T$, with symbol $t(z)=\mathrm{sign}[f(z)]$---this means that $T_{mn}=t_{m-n}$, the $({m-n})$th Fourier coefficient of $t(z)$ \cite{Deift2013}. There is a unitary transformation from the $c_j$ to fermionic modes $d_j$ such that the reduced density matrix for the subsystem $A$ is given by: \begin{align}
\rho_A = \prod_{j=1}^L \frac{1+\nu_j}{2} d^\dagger_j d^{\vphantom\dagger}_j +\frac{1-\nu_j}{2} d^{\vphantom\dagger}_j d^\dagger_j \label{eq:densitymatrix}
\end{align}
(see, for example, Refs.~\cite{Peschel03,Vidal03,Peschel09}). 
Consider the charged moment, where we make the dependence on $L$ explicit: \begin{align}
Z_n(\alpha,L) =\tr (\rho_A^n \rme^{\rmi \alpha \sum_{j=1}^L c^\dagger_jc^{\vphantom\dagger}_j}). \label{eq:chargedmoment}
\end{align}
Using \eqref{eq:densitymatrix}, we then have the following expression for the charged moment:
\begin{align}
Z_n(\alpha,L)= \prod_{j=1}^L \left( \left(\frac{1+\nu_j}{2}\right)^n\rme^{\rmi \alpha}+\left(\frac{1-\nu_j}{2}\right)^n\right).
\end{align}
Note that $Z_n(\alpha,L)$ is periodic in $\alpha$, and so we can restrict without loss of generality to $-\pi<\alpha\leq \pi$. However, $\alpha = \pi$ is a special case that is excluded from our analysis.

Denote by $D_L[t(z)]$ the Toeplitz determinant generated by the symbol $t(z)$. The $\nu_j$ for $1\leq j\leq L$ are eigenvalues of the Toeplitz matrix generated by $\mathrm{sign}[f(z)]$. Hence, they are the zeros of the characteristic polynomial $D_L[\lambda - \mathrm{sign}[f(z)]]$. We will write $t(z,\lambda)=\lambda -  \mathrm{sign}[f(z)]$.
From the residue theorem we have the following expression for the logarithm of the charged moment:
\begin{align}
\log Z_n(\alpha,L)=  \frac{1}{2\pi\rmi} \int_{C} f_n(1+\eps,\lambda,\alpha) \frac{\rmd \log D_L[t(z,\lambda)]}{\rmd\lambda} \rmd \lambda, \label{logchargedmoment}
\end{align}
where 
\begin{align}
f_n(x,\lambda,\alpha) = \log{\left[ \left(\frac{x+\lambda}{2}\right)^n\rme^{\rmi \alpha}+\left(\frac{x-\lambda}{2}\right)^n\right]},
\end{align}
and $C$ is any simple and positively oriented contour containing the interval $[-1,1]$ that does not cross the branch cuts of $f_n(1+\eps,\lambda,\alpha)$. The branch cut structure and the contour of integration will be considered in the following subsection.
 \subsubsection{Branch cuts of the integrand and integration contours for the charged moments.}
 The integrand of equation \eqref{logchargedmoment} is the product of the function $f_n(x,\lambda,\alpha)$ and the logarithmic derivative of a Toeplitz determinant. This determinant is the characteristic polynomial, with zeros corresponding to eigenvalues $\nu_j$ that satisfy $-1\leq \nu_j \leq 1$. This is because they correspond to eigenvalues of the reduced density matrix \eqref{eq:densitymatrix}; hence the logarithmic derivative has singularities on the real axis between $-1\leq \lambda \leq 1$ and nowhere else. 
 
The function $f_n(1+\eps,\lambda,\alpha)$ has branch points at \begin{align}
\frac{1+\eps+\lambda}{1+\eps-\lambda}=\rme^{\rmi\frac{-\alpha+\pi+2\pi m}{n}} = \rme^{\rmi \theta_m} \label{eq:branchpoints}
\end{align}
where the integers $m$ are such that the argument $\theta_m$ satisfies $-\pi< \theta_m\leq\pi$ (this is consistent with the principal logarithm defining $(1+\eps\pm\lambda)^n$ for general $n$). The branch point associated to $\theta_m$ is at $\lambda=\rmi (1+\eps)\tan(\theta_m/2)$: hence, the branch points have zero real part, and the imaginary part is non-zero for $-\pi<\alpha<\pi$. Note that taking the limit $\alpha \rightarrow \pm \pi$, the closest imaginary branch points to the real line approach zero, and in the limit we can no longer set up the problem using the residue theorem.

To fix the contour, we need the imaginary branch points closest to the real line above and below, we will denote them by $\rmi \lambda_\pm$. If there is no such branch point we put $\lambda_\pm = \pm\infty$. Note that the branch point corresponding to $\theta_0$ is above the real axis, and the the branch point corresponding to $\theta_{-1}$ is below. As we vary $n$, there are three different cases that appear:
\begin{enumerate}
\item For $n\geq 2$ there are always at least two solutions to \eqref{eq:branchpoints}:  $\theta_0$ and $\theta_{-1}$. Hence, $\rmi\lambda_+=\rmi (1+\eps)\tan(\frac{\pi-\alpha}{2n})$ and $\rmi\lambda_-=-\rmi(1+\eps) \tan(\frac{\pi+\alpha}{2n})$.
\item For $1\leq n <2$ we have that for $-\pi(n-1)\leq \alpha < \pi(n-1)$, both  $\theta_0$ and $\theta_{-1}$ are solutions to \eqref{eq:branchpoints}. For $\pi(n-1)\leq\alpha<\pi$ we have a single solution $\theta_0$, and for  $-\pi< \alpha< -\pi(n-1)$ we have a single solution $\theta_{-1}$. Depending on the value of $\alpha$ we have, for example,  $\rmi\lambda_+=\rmi (1+\eps)\tan(\frac{\pi-\alpha}{2n})$ or  $\rmi\lambda_+=\rmi \infty$.
\item For $0<n<1$ we have either no solutions or a single solution to \eqref{eq:branchpoints}. For $\pi(1-n)\leq \alpha<\pi$, $\theta_0$ is a solution. For $-\pi<\alpha<-\pi(1-n)$, $\theta_{-1}$ is a solution.
\end{enumerate}
For generic (non-integer) $n$ there are also branch points at $\lambda=\pm(1+\eps)$.

 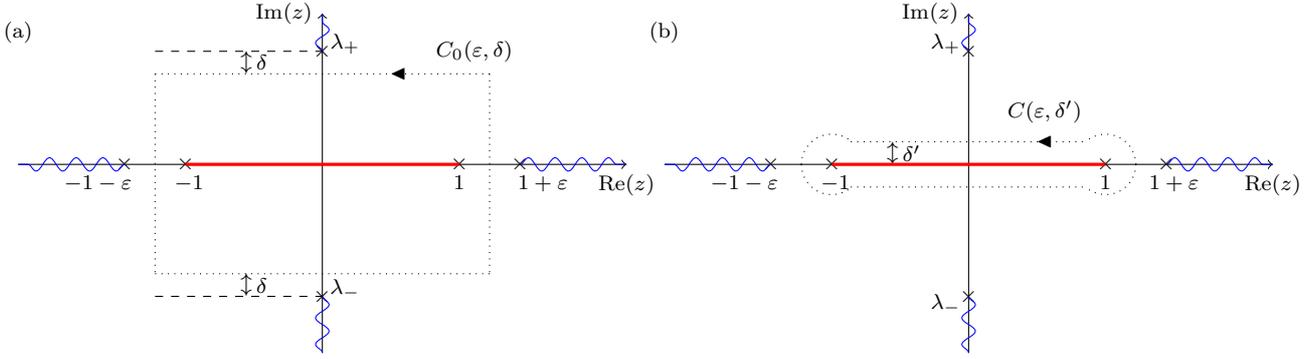
\begin{figure}
 \begin{tikzpicture}
 \draw[font=\scriptsize]  (-4,1.75) node {(a)};
  \draw[font=\scriptsize]  (-1,1.35) node {$\updownarrow$} node [right] {$\delta$};
    \draw[font=\scriptsize]  (-1,-1.6) node {$\updownarrow$} node [right] {$\delta$};
  \draw[font=\scriptsize]  (2,1.5) node {$C_0(\eps,\delta)$};
 \draw[font=\scriptsize] [->] (-4,0) -- (4,0) node [below ]  {$\Re(z)$};
    \draw[font=\scriptsize] [->] (0,-2.5) -- (0,2) node [left] {$\Im(z)$};
\draw[font=\scriptsize] (0.3,1.6) node {$\lambda_+$};\draw[font=\scriptsize] (0,1.5) node {$\times$};
\draw[font=\scriptsize] (0.3,-1.65) node {$\lambda_-$};\draw[font=\scriptsize] (0,-1.75) node {$\times$};
\draw[font=\scriptsize] (2.6,0) node {$\times$};
\draw[font=\scriptsize]  (-2.6,0) node {$\times$};
\draw[font=\scriptsize]  (1.8,0) node {$\times$};
\draw [font=\scriptsize] (-1.8,0) node {$\times$};
\draw[font=\scriptsize]  (2.9,-0.25) node {$1+\eps$};
\draw[font=\scriptsize]  (1.8,-0.25) node {$1\vphantom{+\eps}$};
\draw[font=\scriptsize]  (-2.95,-0.25) node {$-1-\eps$};
\draw[font=\scriptsize]  (-1.75,-0.25) node {$-1\vphantom{-\eps}$};
  \path [draw=blue,snake it]  (2.6,0) -- (4,0);
  \path [draw=blue,snake it]  (-2.6,0) -- (-4,0);
  \path [draw=blue,snake it]  (0,1.5) -- (0,2);
  \path [draw=blue,snake it]  (0,-1.75) -- (0,-2.5);
  \draw [red, very thick] (-1.8,0) -- (1.8,0);
   \draw [dashed] (-2.2,1.5) -- (0,1.5);
   \draw [dashed] (-2.2,-1.75) -- (0,-1.75);
 \draw [dotted] (-2.2,1.2) -- (2.2,1.2);
  \draw [dotted] (-2.2,-1.45) -- (2.2,-1.45);
 \draw [dotted] (2.2,1.2) -- (2.2,-1.45);
 \draw [dotted] (-2.2,1.2) -- (-2.2,-1.45);
 \draw (1,1.2) node [font=\scriptsize,rotate = 90]  {$\blacktriangle$};
\begin{scope}[shift={(8.5,0)}]
 \draw[font=\scriptsize]  (-4,1.75) node {(b)};
   \draw[font=\scriptsize]  (1,.7) node {$C(\eps,\delta')$};
    \draw[font=\scriptsize]  (-1,0.15) node {$\updownarrow$} node [right] {$\delta'$};
 \draw[font=\scriptsize] [->] (-4,0) -- (4,0) node [below ]  {$\Re(z)$};
    \draw[font=\scriptsize] [->] (0,-2.5) -- (0,2) node [left] {$\Im(z)$};
\draw[font=\scriptsize] (-0.3,1.6) node {$\lambda_+$};\draw[font=\scriptsize] (0,1.5) node {$\times$};
\draw[font=\scriptsize] (-0.3,-1.85) node {$\lambda_-$};\draw[font=\scriptsize] (0,-1.75) node {$\times$};
\draw[font=\scriptsize] (2.6,0) node {$\times$};
\draw[font=\scriptsize]  (-2.6,0) node {$\times$};
\draw[font=\scriptsize]  (1.8,0) node {$\times$};
\draw [font=\scriptsize] (-1.8,0) node {$\times$};
\draw[font=\scriptsize]  (2.7,-0.25) node {$1+\eps$};
\draw[font=\scriptsize]  (1.8,-0.25) node {$1\vphantom{+\eps}$};
\draw[font=\scriptsize]  (-2.95,-0.25) node {$-1-\eps$};
\draw[font=\scriptsize]  (-1.75,-0.25) node {$-1\vphantom{-\eps}$};
  \path [draw=blue,snake it]  (2.6,0) -- (4,0);
  \path [draw=blue,snake it]  (-2.6,0) -- (-4,0);
  \path [draw=blue,snake it]  (0,1.5) -- (0,2);
  \path [draw=blue,snake it]  (0,-1.75) -- (0,-2.5);
  \draw [red, very thick] (-1.8,0) -- (1.8,0);
 \draw [dotted] (-1.55,0.3) -- (1.55,.3);
  \draw [dotted] (-1.55,-.3) -- (1.55,-.3);
\draw[dotted] (2.2,0) arc (0:130:0.4);
\draw[dotted] (2.2,0) arc (0:-130:0.4);
\draw[dotted] (-2.2,0) arc (180:50:0.4);
\draw[dotted] (-2.2,0) arc (180:310:0.4);
 \draw (1,.3) node  [font=\scriptsize,rotate = 90] {$\blacktriangle$};
\end{scope}
\end{tikzpicture}
\caption{Contours of integration for calculating the logarithm of the charged moment. The bold (red) line contains singularities of the logarithmic derivative of $D_L[t(z,\lambda)]$, while wavy (blue) lines correspond to branch cuts of $f_n(1+\eps,\lambda,\alpha)$. $C_0(\eps,\delta)$ is the initial contour, where we take the asymptotic expansion of the Toeplitz determinant. Then we deform the integral to $C(\eps,\delta')$ to evaluate the terms in this expansion. We have branch points at $\rmi\lambda_+ = \rmi (1+\eps)\tan(\frac{\pi-\alpha}{2n})$ and $\rmi\lambda_- =-\rmi(1+\eps) \tan(\frac{\pi+\alpha}{2n}) $, and hence assume that $-\pi<\alpha<\pi$ and $n\geq2$. The case where $n<2$ is described in the text.}
\label{fig:contours}
 \end{figure}
 
 We will use two contours of integration, sketched in Fig.~\ref{fig:contours}. The first contour, $C_0(\eps,\delta)$, is a rectangular contour with corners, for finite\footnote{If $\lambda_+$ (or $\lambda_-$) is infinite we take a rectangle with corner at $\lambda =1+\eps/2 + \rmi R$ (or at $\lambda = -1-\eps/2- \rmi R$), and then take the limit as $R\rightarrow \infty$. } $\lambda_\pm$, at $\lambda = 1+\eps/2 + \rmi( \lambda_+ -\delta)$ and $\lambda = -1-\eps/2 + \rmi( \lambda_- +\delta)$. This contour will be useful for defining the asymptotic expansion with an explicit and uniform estimate for the error. $C_0(\eps,\delta)$ may be deformed to $C(\eps,\delta')$. Below we will take the limit as $\eps,\delta' \rightarrow 0$ and the contour integral reduces to two integrals infinitesimally above and below the red cut---these are the integrals appearing in related calculations in Refs.~\cite{Jin2004,Keating2004,Bonsignori19,Fraenkel20}.
 
\subsubsection{Fisher-Hartwig asymptotics for the Toeplitz determinant}\label{sec:asymptotics}
Now, to find the large $L$ asymptotics of $\log Z_n(\alpha)$ we need the large $L$ asymptotics of the Toeplitz determinant $\log D_L[t(z,\lambda)]$. This is a Toeplitz determinant with Fisher-Hartwig jump singularities, and the asymptotics may be rigorously evaluated using standard results \cite{Basor79,Deift2014}, our treatment closely follows Refs.~\cite{Jin2004,Keating2004}. 
First of all, one needs to put the symbol in a canonical form:
\begin{align}
t(z)= \rme^{V(z)} z^{\sum_r \beta_r} \prod_{r} g_{z_r,\beta_r(\lambda)}(z) z_r^{-\beta_r(\lambda)} \label{eq:canonical}
\end{align}
where 
\begin{align}
g_{z_j,\beta_j}(z) = \begin{cases}
\rme^{\rmi \pi \beta_j} \qquad &0 \leq \arg(z) <\theta_j\\
\rme^{-\rmi \pi \beta_j} \qquad &\theta_j \leq \arg(z) <2\pi
\end{cases} \qquad \textrm{for~} z_j=\rme^{\rmi\theta_j}.
\end{align}
Now, suppose that $f(z)$ has $2N$ zeros at\footnote{We put all arguments between $0$ and $2\pi$ for consistency with Ref. \cite{Deift2014}.} $\{z_1=\rme^{\rmi k_1}, \dots,z_N=\rme^{\rmi k_N}, z_{N+1} =\rme^{\rmi(2\pi-k_N)},\dots, z_{2N}= \rme^{\rmi(2\pi-k_1)}\}$. Then we can have a canonical form for $t(z,\lambda)$ with the following choices\footnote{$V_k$ is the $k$th Fourier coefficient of $V(z)$.}:
\begin{align}
\rme^{V(z,\lambda)}&= \rme^{V_0(\lambda)}=  (\lambda+1) \left(\frac{\lambda+1}{\lambda-1}\right)^{\sum_{r=1}^N (-1)^{N+1-r}k_r/\pi}\\
\beta_r &= (-1)^{N+1-r} \beta(\lambda)  \qquad\qquad  1\leq r \leq 2N \label{eq:beta}
\end{align}
where
\begin{align}
\beta(\lambda) &= \frac{1}{2\pi\rmi} \log\left(\frac{\lambda+1}{\lambda-1}\right); \qquad\qquad -\pi \leq \arg\left(\frac{\lambda+1}{\lambda-1} \right) <\pi. \label{eq:lambdabranch}
\end{align}
This representation implicitly assumes that $f(\rme^{\rmi \pi})<0$; this means that the sign in \eqref{eq:standardform} is equal to $-1$. It suffices to consider only that case: suppose we transform $f(z) \rightarrow -f(z)$, then the eigenvalues of the Toeplitz matrix generated by $\mathrm{sign}[f(z)]$ go from $\nu_j \rightarrow -\nu_j$. Hence:
\begin{align}
Z_n(\alpha,L)\Big\vert_{-f(z)} =  \prod_{j=1}^L \left( \left(\frac{1-\nu_j}{2}\right)^n\rme^{\rmi \alpha}+\left(\frac{1+\nu_j}{2}\right)^n\right) = \rme^{\rmi\alpha L}Z_n(-\alpha,L)\Big\vert_{f(z)}. \label{eq:-f(z)}
\end{align} 
Thus if $f(\rme^{\rmi \pi})>0$, we can find the charged moment by calculating the charged moment of $-f(z)$ using the representation above.

Taking the result from \cite{Deift2014} we then have the following asymptotics:
\begin{align}
\log D_L[t(z,\lambda)] = V_0(\lambda) L - 2N\beta(\lambda)^2 \log L + \log E(\lambda) +o(1) 
\end{align}
 where 
\begin{align}
E(\lambda) = \bigg(\sigma^{- \beta(\lambda)^2}G(1+\beta(\lambda))G(1-\beta(\lambda)) \bigg)^{2N};\label{eq:constant}
\end{align}
recall that $\sigma$ was defined in \eqref{eq:sigma}. 

In terms of the length scale $L_\sigma = \sigma L$, we can rewrite this as:
\begin{align}
\log D_L[t(z,\lambda)] = V_0(\lambda) L - 2N\beta(\lambda)^2 \log L_\sigma + 2N \log\Bigg(G(1+\beta(\lambda))G(1-\beta(\lambda))\Bigg)  +o(1) .
\end{align}

From Ref.~\cite{Deift2014}, the error term is $o(1) = O(L^{\lvert\lvert\beta(\lambda) \rvert\rvert -1})$, where $\lvert\lvert\beta \rvert\rvert$ is given by  $\max_{j,k}  \lvert\Re(\beta_j) - \Re(\beta_k)\rvert$. This error term is uniform for compact sets of $\beta$ such that $\lvert\lvert\beta \rvert\rvert<1$, and is analysed in Appendix \ref{app:errorterm} for the contour of integration $C_0(\eps,\delta)$. There we show that we have:
\begin{align}
\lvert\lvert \beta(\lambda) \rvert\rvert -1 \leq \underbrace{\max\Bigg\{-\frac{1}{2},  -\frac{1}{n} \left(1-\frac{\lvert\alpha\rvert}{\pi}\right) \Bigg\}}_{-\mu(n,\alpha)} +O(\delta,\eps). \label{eq:beta1}
\end{align}
We can take the limit $\eps,\delta\rightarrow 0$ and, for a fixed $-\pi<\alpha<\pi$ and $n>0$, we can hence uniformly bound the next-to-leading order term in the asymptotics. 
\subsubsection{Asymptotics of the charged moment: Proof of Result \ref{resultone}}

In order to evaluate the charged moment we insert these asymptotics into the integral and then deform the contour from $C_0(\eps,\delta)$ to $C(\eps,\delta')$. I.e., we have:
\begin{align}
\log Z_n(\alpha,L)&= \lim_{\eps,\delta\rightarrow 0}\frac{1}{2\pi\rmi} \int_{C_0(\eps,\delta)} f_n(1+\eps,\lambda,\alpha) \frac{\rmd \log D_L[t(z,\lambda)]}{\rmd\lambda} \rmd \lambda\nonumber\\
&=  \lim_{\eps,\delta\rightarrow 0}\frac{\rmi}{2\pi} \int_{C_0(\eps,\delta)}  \frac{\rmd f_n(1+\eps,\lambda,\alpha)}{\rmd\lambda} \log D_L[t(z,\lambda)] \rmd \lambda\nonumber\\
&=  a_0 L + a_1\log L_\sigma +a_2 + O(L^{-\mu(n,\alpha)}),\label{eq:contourint}
\end{align}
where 
\begin{align}
a_0 &=\lim_{\eps,\delta'\rightarrow 0}\frac{1}{2\pi\rmi} \int_{C(\eps,\delta')} f_n(1+\eps,\lambda,\alpha) \Big(\frac{1+\sum_{r=1}^N (-1)^{N+1-r} k_r/\pi}{1+\lambda} + \frac{\sum_{r=1}^N (-1)^{N+1-r}  k_r/\pi}{1-\lambda} \Big)\rmd \lambda\\
a_1 &= N\lim_{\eps,\delta'\rightarrow 0}\frac{1}{2\pi\rmi} \int_{C(\eps,\delta')} f_n(1+\eps,\lambda,\alpha) \frac{\rmd \big(-2\beta(\lambda)^2\big)}{\rmd \lambda} \rmd \lambda = \frac{2N}{\pi^2} \lim_{\eps,\delta'\rightarrow 0}\int_{C(\eps,\delta')} f_n(1+\eps,\lambda,\alpha) \frac{\beta(\lambda)}{1-\lambda^2} \rmd \lambda  \label{eq:a1}\\
a_2 &= \frac{N}{\pi\rmi}  \lim_{\eps,\delta'\rightarrow 0}\int_{C(\eps,\delta')} f_n(1+\eps,\lambda,\alpha) \frac{\rmd \log \big(G(1+\beta(\lambda))G(1-\beta(\lambda))\big)}{\rmd \lambda} \rmd \lambda \label{eq:a2} .
\end{align}
Note that we integrate by parts initially to avoid differentiating the asymptotic expansion---for each individual integral we then integrate by parts again.
These integrals can be evaluated as in Refs.~\cite{Bonsignori19,Fraenkel20}. In particular we have:
\begin{align}
a_0 &= -\rmi \alpha \sum_{r=1}^N (-1)^{N+1-r} k_r/\pi\\
a_1 &= -N\Bigg( \frac{1}{6}\left(n-\frac{1}{n}\right)+\frac{2}{n} \left(\frac{\alpha}{2\pi}\right)^2\Bigg)\\
a_2 &=N ~ \Upsilon(n,\alpha)
\end{align}
where $ \Upsilon(n,\alpha)$ is defined in \eqref{def:upsilon}; for further details see the next subsection. 

Now, recall that in our analysis we assumed that $f(\rme^{\rmi \pi})<0$; in this case the mean density of charges in the ground state is given by $\langle Q_A\rangle/L =\sum_{r=1}^N (-1)^{N-r} k_r/\pi$. In the case that $f(\rme^{\rmi \pi})>0$ we have that $\langle Q_A\rangle/L =(\pi - \sum_{r=1}^N (-1)^{N-r} k_r)/\pi $ . Thus using  \eqref{eq:-f(z)}, and noting that the other terms are even in $\alpha$, we have Result \ref{resultone} for all cases. Note that the dependence on $\langle Q_A\rangle$ is as expected \cite{Bonsignori19}, since from \eqref{eq:chargedmoment} we have that $Z_1(\alpha,L) = 1+ \rmi \alpha \langle Q_A\rangle + O(\alpha^2)$. This agrees with the calculations above: fixing $n=1$ gives $Z_1(\alpha,L) = \rme^{  \rmi \alpha \langle Q_A\rangle + O(\alpha^2)}$. To see this, we use $a_1 = -2N \left(\frac{\alpha}{2\pi}\right)^2$ and from \eqref{def:upsilonexpansion} we have $a_2 = O(\alpha^2)$ (see also the following subsection). 
\subsection{Analysis of \texorpdfstring{$\Upsilon(n,\alpha)$}{Y(n,a)}}\label{sec:upsilon}
By evaluating the integral \eqref{eq:a2}, we have:
\begin{align}
 \Upsilon(n,\alpha)&= \rmi n \int_{-\infty}^\infty \left(\tanh(\pi w) - \tanh(n \pi w +\rmi \alpha/2) \right)\log \frac{\Gamma(1/2+\rmi w)}{\Gamma(1/2-\rmi w)} \rmd w\label{eq:upsilon1}\\
 &=-\frac{1}{\pi^2} \int_0^\infty \int_0^\infty \log\Bigg(\frac{2\cos(\alpha) +2\cosh(n u)}{\left(2\cosh(\frac{u}{2})\right)^{2n}}\Bigg)\Bigg(\frac{\rme^{-t}}{t}-\frac{\cos(\frac{ut}{2\pi})}{2\sinh(t/2)}\Bigg)\rmd t\rmd u \label{eq:upsilon2}.
\end{align}
The derivation of $ \Upsilon(n,\alpha)$ follows the same steps as for the corresponding term $\Upsilon(n)=\Upsilon(n,0)$ given for the XX model in Ref.~\cite{Jin2004}. Hence, the expressions agree in that case; we refer the reader to Refs.~\cite{Bonsignori19,Fraenkel20} for details. The form given in \eqref{eq:upsilon2} (for a derivation see \cite{Fraenkel20}) makes the following features explicit: $\Upsilon(n,\alpha)$ is real valued and is an even function of $\alpha$. In the expression \eqref{eq:upsilon1}, we can use the Taylor expansion of $\tanh(n \pi w +\rmi \alpha/2)$ about $\alpha=0$, which converges uniformly in $w$ for $\lvert \alpha \rvert <\pi$. Hence, using the notation of \cite{Bonsignori19}, we have the series expansion: 
\begin{align}
 \Upsilon(n,\alpha)&= \Upsilon(n) + \alpha^2 \gamma_2(n)+ \alpha^4 \gamma_4(n) + \epsilon(n,\alpha) 
\end{align}
where $\epsilon(n,\alpha) = O(\alpha^6)$ and $\gamma_2(n)$ and $\gamma_4(n)$ are given in Eq.~\eqref{eq:gamma2,4}. This expansion will be used in calculating asymptotics of the symmetry-resolved moments.

An alternative formula for $\Upsilon(n,\alpha)$ can be derived by integrating \eqref{eq:a2} by parts leading to:
\begin{align}
N\Upsilon(n,\alpha) &= -\frac{N}{\pi\rmi}  \lim_{\eps,\delta'\rightarrow 0}\int_{C(\eps,\delta')}  \frac{\rmd}{\rmd \lambda} \Big(f_n(1+\eps,\lambda,\alpha)\Big)\log\big( G(1+\beta(\lambda))G(1-\beta(\lambda))\big) \rmd \lambda.
\end{align}
The function $\rmd f_n(1+\eps,\lambda,\alpha) /\rmd\lambda$ has simple poles at $\lambda_m = \rmi(1+\eps) \tan(\theta_m/2)$, where the $\theta_m$ are defined in \eqref{eq:branchpoints}.  For $n\in\mathbb{Z}_+$ these are the only singularities and deforming the contour to infinity, the residue theorem leads to:
\begin{align}
\Upsilon(n,\alpha) &=2\sum_{m=0}^{n-1} \log\Bigg( G\Big(1-\frac{\alpha}{2\pi n}+\frac{2m+1-n}{2n} \Big) G\Big(1+\frac{\alpha}{2\pi n}+ \frac{2m+1-n}{2n}  \Big)\Bigg). \label{eq:upsilonresidues}\end{align}
Note that this form is also manifestly symmetric in $\alpha$. This result was previously derived using residue calculus in Ref.~\cite{Estienne21} and coincides with the result of lattice form-factor calculations described in Section \ref{sec:formfactor}. 

We have the following expansion for the summands in the above expression:
\begin{align}
 \log\Bigg( G\Big(x-\frac{\alpha}{2\pi n}\Big) G\Big(x+\frac{\alpha}{2\pi n}\Big)\Bigg) =& \log(G(x)^2) +\frac{1}{4n^2\pi^2} \left(-1+\psi(x)+(x-1)\psi^{(1)}(x) \right)\alpha^2\nonumber\\
 &\qquad + \frac{1}{192n^4\pi^4} \left(3\psi^{(2)}(x)+(x-1)\psi^{(3)}(x) \right)\alpha^4+O(\alpha^6),
\end{align}
(recall that $\psi(x)$ is the usual polygamma function). Using also that $G(1)=1$, we have the result given in \eqref{eq:gamma2,4n} for $n\in \mathbb{Z}_+$. Using this formula for $\gamma_2(1)$ we then immediately have the known result $\gamma_2(1) = -\frac{1+\gamma_E}{2\pi^2}$ \cite{Xavier18,Bonsignori19,Fraenkel20}; we also have that $\gamma_4(1) = \frac{\psi^{(2)}(1)}{32\pi^4}$. More generally, these expressions give a closed formula that avoids the need for numerical integration to evaluate $\gamma_2(n)$ and $\gamma_4(n)$ for integer $n$. The above method is straightforwardly extended to find higher terms in the expansion of $\Upsilon(n,\alpha)$ by expanding \eqref{eq:upsilonresidues}.

For general $0<n\in\mathbb{R}$, the function $\rmd f_n(1+\eps,\lambda,\alpha) /\rmd\lambda$ has branch points at $\lambda = \pm (1+\eps)$; and we take the branch cuts along the real line. Deforming the contour to infinity we have additional contributions from integrating along the branch cuts, as well as residues. Defining $m_1=\lfloor n/2-1/2+\frac{\alpha}{2\pi}\rfloor$ and $m_2 = \lfloor n/2+1/2-\frac{\alpha}{2\pi}\rfloor$, a calculation leads to:
\begin{align}
&\Upsilon(n,\alpha) =
2\sum_{m=0}^{m_1} \log\Bigg( G\Big(1 +  \frac{1}{2 n}\left(2m+1-n -\frac{\alpha}{\pi} \right)\Big) G\Big(1- \frac{1}{2 n}\left(2m+1-n -\frac{\alpha}{\pi} \right)  \Big)\Bigg)\nonumber\\
&+2\sum_{m=1}^{m_2} \log\Bigg( G\Big(1 +  \frac{1}{2 n}\left(1+n-2m -\frac{\alpha}{\pi} \right)\Big) G\Big(1- \frac{1}{2 n}\left(1+n-2m -\frac{\alpha}{\pi} \right)  \Big)\Bigg)+ \frac{8n \sin(n \pi)}{\pi}I(n,\alpha),
\end{align}
where the branch cut contribution corresponds to:
\begin{align}
I(n,\alpha) =\int_0^\infty &\frac{x^{n-1} (x+2)^{n-1} \left(\cos (\alpha) \left(x^{2 n}+(x+2)^{2 n}\right)+2 x^n (x+2)^n \cos (\pi  n)\right)}{\Big\lvert (x+2)^{2n}+ 2x^n (x+2)^{n}\rme^{\rmi n\pi}\cos(\alpha) + \rme^{2\rmi n\pi} x^{2n} \Big\rvert^2}\nonumber\\
&\times \log\Bigg(G\left(1+\frac{ 1}{2\pi \rmi}\log(1+2/x)\right)G\left(1-\frac{ 1}{2\pi \rmi}\log(1+2/x)\right) \Bigg) \rmd x.
\end{align}

\subsection{Symmetry-resolved moments}\label{sec:prooftwo}

We have that
\begin{align}
\mathcal{Z}_n(q,L) = \frac{1}{2\pi} \int_{-\pi}^\pi \rme^{-\rmi q\alpha} Z_n(\alpha,L) \rmd \alpha.
\end{align}
We are interested in the case that $q_\Delta=q-\langle Q_A\rangle $ is $O(1)$ for large $L$. A corresponding analysis can be made for $q_\Delta$ at different scales. We have from above that: 
\begin{align}
&Z_n(0,L) = L_\sigma^{-\frac{N}{6}\left(n-\frac{1}{n}\right)}\rme^{N \Upsilon(n)}\left(1+O\left(L^{\max\left\{-\frac{1}{2},-\frac{1}{n}\right\}}\right)\right).  \end{align}

With these definitions, we have:
\begin{align}
\mathcal{Z}_n( q) &= \frac{Z_n(0,L)}{2\pi} \int_{-\pi}^\pi  \rme^{-\rmi  q_\Delta \alpha- \frac{\alpha^2}{2n\pi^2} N\log(  L_\sigma) + N\left(\alpha^2 \gamma_2(n) +  \alpha^4 \gamma_4(n)+\epsilon(n,\alpha)\right)} \left(1+O(L^{-\mu(n,\alpha)}) \right)\rmd \alpha .
\end{align}
Then let us define:
\begin{align}
a(n,L)
&=\frac{N}{2n\pi^2}\left( \log(  L_\sigma) -2n\pi^2\gamma_2(n) \right)= O(\log L) . \end{align}
Note that $a(n,L)$ is a real-valued function, and for $L$ sufficiently large it will be positive-valued.
We will split the integral over $\alpha$ into three pieces:
\begin{align}
\mathcal{Z}_n( q) &=\nonumber \underbrace{\frac{1}{2\pi} \int_{-\pi}^{-\pi/2} \rme^{-\rmi (q_\Delta +\langle Q_A\rangle )\alpha} Z_n(\alpha,L) \rmd \alpha}_{I_1}+\underbrace{\frac{1}{2\pi} \int_{-\pi/2}^{\pi/2} \rme^{-\rmi (q_\Delta +\langle Q_A\rangle )\alpha} Z_n(\alpha,L) \rmd \alpha}_{I_2}\\ &\quad+\underbrace{\frac{1}{2\pi} \int_{\pi/2}^{\pi} \rme^{-\rmi (q_\Delta +\langle Q_A\rangle )\alpha} Z_n(\alpha,L) \rmd \alpha}_{I_3}.
\end{align}
Asymptotically $I_2$ dominates $I_1$ and $I_3$ as we will now show. First, let us analyse $I_2$ as $L\rightarrow \infty$. Note that for $\lvert \alpha \rvert\leq \pi/2$ we have $-\mu(n,\alpha) \leq \max\{-\frac{1}{2},-\frac{1}{2n}\}$. Then:
\begin{align}
I_2 &= \frac{Z_n(0,L)}{2\pi} \int_{-\pi/2}^{\pi/2}  \rme^{ -\alpha^2 a(n,L)-\rmi  q_\Delta \alpha +  N\alpha^4 \gamma_4(n)+N\epsilon(n,\alpha)} \left(1+O\left(L^{\max\left\{-\frac{1}{2},-\frac{1}{2n}\right\}}\right)\right)\rmd \alpha 
\nonumber\\&= \frac{1}{\sqrt{a(n,L)}}\frac{Z_n(0,L)}{2\pi} \int_{-\pi \sqrt{a(n,L)}/2}^{\pi \sqrt{a(n,L)}/2}  \exp\Bigg( -\alpha'^2 -\rmi \frac{ q_\Delta}{\sqrt{a(n,L)}} \alpha' \Bigg) \Bigg(1+\frac{N \alpha'^4}{a(n,L)^2} \gamma_4(n) +O(\log(L)^{-3})\Bigg) \rmd \alpha' \nonumber\\&= \frac{Z_n(0,L)}{\sqrt{4 \pi a(n,L)}} \exp\left(- \frac{ q_\Delta^2}{4a(n,L)}\right)  \Bigg(1+\frac{3}{4}\frac{N \gamma_4(n)}{a(n,L)^2}+ O(\log{(L)}^{-3}) +e(L) \Bigg).
\end{align}
In the last line we extend the limits of integration to infinity. The corresponding error function asymptotics result in a subdominant error term 
\begin{align}
e(L)=O\left({ \exp\left(\frac{ q_\Delta^2}{4a(n,L)}\right) L^{-\frac{N}{8n}}}(\log (L))^{-1/2}\right)\end{align} that we can ignore\footnote{Note that the power appearing here depends on the arbitrary choice of $\alpha = \pm\pi/2$ that we use to cut off the integral away from $\alpha =\pm \pi$. We are free to choose any $\alpha = \pm (\pi - x)$ for fixed $x>0$ and this will lead to a different algebraic decay. This is always subdominant to the expansion in negative powers of $\log(L)$ that comes from expanding $\exp(N \gamma_4(n) + N\epsilon(n,\alpha))$ and doing the Gaussian integral.}. For $I_3$ we have the bound:
\begin{align}
\lvert I_3 \rvert &= \Bigg \lvert \frac{Z_n(0,L)}{2\pi} \int_{\pi/2}^{\pi}  \rme^{ -\alpha^2 a(n,L)-\rmi  q_\Delta \alpha +N  \alpha^4 \gamma_4(n)+N\epsilon(n,\alpha)} \left(1+O(1)\right)\rmd \alpha  \Bigg\rvert
\nonumber\\&\leq \frac{\mathrm{const}}{\sqrt{a(n,L)}}\frac{Z_n(0,L)}{2\pi} \int_{\pi \sqrt{a(n,L)}/2}^{\pi \sqrt{a(n,L)}}  \exp( -\alpha'^2 ) \left(1+O(\log(L)^{-2})\right) \rmd \alpha' = O\left(Z_n(0,L){L^{-\frac{N}{8n}}}(\log (L))^{-{1}}\right).
\end{align}
Similarly $\lvert I_1 \rvert = O\left(Z_n(0,L){L^{-\frac{N}{8n}}}(\log (L))^{-{1}}\right)$. These contributions are subdominant to the expansion in negative powers of $\log(L)$ that appear in $I_2$. We hence have the large $L$ asymptotics claimed in Result \ref{resulttwo}:
\begin{align}
\mathcal{Z}_n( q) &= \frac{Z_n(0,L)}{\sqrt{4\pi a(n,L)}} \exp\left(- \frac{ q_\Delta^2}{4a(n,L)}\right) \Bigg(1+\frac{3}{4}\frac{N\gamma_4(n)}{a(n,L)^2}+ O(\log{(L)}^{-3})\Bigg).\label{eq:mathcalZ}
\end{align}

\subsection{Symmetry-resolved entropies}\label{sec:proofthree}
The symmetry-resolved R\'enyi entropies are defined by:
\begin{align}
S_n(q) &= \frac{1}{1-n}\Bigg( \log\left(\mathcal{Z}_n( q)\right) - n \log\left( \mathcal{Z}_1( q)\right) \Bigg).
\end{align}
From \eqref{eq:mathcalZ}, it is then straightforward to extract the asymptotics. First:

\begin{align}
S_n(q_\Delta) &= S_n - \frac{1}{2(1-n)}\log\left(4\pi a(n,L)\right)+ \frac{n}{2(1-n)}\log{\left(4\pi a(1,L)\right)} + \frac{1}{1-n}\left(- \frac{ q_\Delta^2}{4a(n,L)}+\frac{n q_\Delta^2}{4a(1,L)}\right) \nonumber\\
&\qquad + \frac{1}{1-n} \log\Bigg(1+\frac{3}{4}\frac{N\gamma_4(n)}{a(n,L)^2}+ O(\log{(L)}^{-3})\Bigg) - \frac{n}{1-n}\log\Bigg(1+\frac{3}{4}\frac{N\gamma_4(1)}{a(1,L)^2}+ O(\log{(L)}^{-3})\Bigg);
\end{align}
where $S_n$ is the usual R\'enyi entropy\footnote{As emphasised in \cite{Bonsignori19} this is the exact value of $S_n$, we then proceed to evaluate it asymptotically.}, with asymptotics:
\begin{align}
S_n =  \Bigg(\frac{N}{6}\left(1+\frac{1}{n}\right)\log( L_\sigma)  + \frac{N}{1-n} \Upsilon (n,0) \Bigg)+O(L^{\max\{-1/2,-1/n\}}).
\end{align}
Expanding the other terms we have: 
\begin{align}
S_n(q) &= S_n- \frac{1}{2(1-n)}\log\Big(4\pi a(n,L)\Big)+ \frac{n}{2(1-n)}\log{\Big(4\pi a(1,L)\Big)}\nonumber\\&+\frac{1}{1-n}\left(- \frac{ q_\Delta^2}{4a(n,L)}+\frac{n q_\Delta^2}{4a(1,L)}\right) + \frac{3}{4(1-n)}\Bigg(\frac{N\gamma_4(n)}{a(n,L)^2} - n\frac{ N\gamma_4(1)}{a(1,L)^2}\Bigg)+ O(\log{(L)}^{-3}).\end{align}
Then inserting the definition of $a(n,L)$, writing $\log(L_\sigma) =  \log(L)+\log(\sigma)$, and grouping powers of $\log(L)$ leads to Result \ref{resultthree}. 

  \section{Discussion}\label{sec:discussion}
  We now discuss our results and put them in context. We focus on Result \ref{resultone}, since given that, the others follow. First in Section \ref{sec:CFT} we explain the calculation of $Z_n(\alpha)$ (we will drop the explicit $L$ dependence in this section) using the twist-field method. We show how this recovers the scaling dimension as well as the oscillatory factor in Result \ref{resultone}, as well as allowing us to derive the same properties of the subdominant terms. Then in Section \ref{sec:formfactor}, we discuss another approach to calculating the exact leading term by identifying the coefficients of the CFT expansion with certain twisted overlaps in the lattice model. As with the CFT method, this applies more generally to models that are not free-fermion, but we see for the free-fermion case that this approach recovers the exact formula. Section \ref{sec:connection} contains more details of the connections between our approach and the CFT and form factor approaches. We also show how, under certain assumptions, one could reconstruct Result \ref{resultone} from the CFT analysis along with known lattice results. In Section \ref{sec:decoupled} we study a simple limiting case, a stack of decoupled chains. Finally in Section \ref{sec:FH} we give a discussion of subleading terms from Toeplitz determinant theory and the connection to the CFT expansion.
  \subsection{Field theory approach}\label{sec:CFT}
The usual method for calculating the R\'enyi entropy $S_n(L)$ in CFT, for integer $n>1$, is to use the replica trick \cite{Holzhey94,Calabrese04,Calabrese09}. This leads to studying the CFT on an $n$-sheeted Riemann surface, that in turn corresponds to a $\mathbb{Z}_n$--orbifold CFT on the complex plane. The moments $\tr(\rho_A^n)$ then correspond to correlators of certain branch-point twist operators $\mathcal{T}_n$. Following Refs.~\cite{Goldstein18,Xavier18,Bonsignori19,Estienne21}, in order to calculate the charged moments, we dress the branch-point twist operator with another operator. This is the CFT field $\mathcal{V}_\alpha$ corresponding to the lattice operator $\rme^{\rmi \alpha Q_A}$; the resulting composite twist operator is $\mathcal{T}_{n,\alpha}=\mathcal{T}_{n}\mathcal{V}_\alpha$. Denoting the corresponding antitwist operator by $\tilde{\mathcal{T}}$, and taking the subsystem $A$ to be an interval of length $L$, we have:
\begin{align}
Z_n(\alpha) = \langle \mathcal{T}_{n,\alpha}(L) \tilde{\mathcal{T}}_{n,\alpha}(0)  \label{eq:CFTformula}\rangle.
\end{align}
The scaling behaviour of this correlator can be derived \cite{Goldstein18}: assuming that $\mathcal{V}_\alpha$ is a primary operator with dimensions $h_\alpha$, $\overline{h}_\alpha$, then the $\mathcal{T}_{n,\alpha}$ behave as primary operators with scaling dimension:
\begin{align}
h_{n,\alpha} = \frac{c}{24}(n-1/n) + h_\alpha/n \qquad \overline{h}_{n,\alpha} = \frac{c}{24}(n-1/n) + \overline{h}_\alpha/n, \label{eq:scalingdim}
\end{align}
where $c$ is the central charge. For further details see the discussion in Refs. \cite{Goldstein18,Xavier18,Bonsignori19,Estienne21}. 

The previous result is rather general. Let us now turn our attention to the family of models considered in this paper, and in particular let us focus on the case where all Fermi velocities are equal so that we have a low energy CFT (we explain below how to think about other cases). Then we can use the general result taking $c=N$ and, considering the theory of $N$ bosonised complex fermions, $\mathcal{V}_\alpha$ is a particular vertex operator with $h_\alpha = \overline{h}_\alpha = \frac{N}{2} \left(\frac{\alpha}{2\pi}\right)^2$. We justify this in the following section.

\subsubsection{The field $\mathcal{V}_\alpha$}
To identify $\mathcal{V}_\alpha$ we give a brief treatment of the field content of this theory; further details may be found in references \cite{Affleck88,Ginsparg90,Fisher97,Polchinski98,diFrancesco99,Sachdev01}. We will use a field-theoretic bosonisation, ignoring certain subtleties, such as Klein factors. The constructive approach \cite{vonDelft1998} can be clearer, but our main aim here is to outline the solution and show that it agrees with the rigorous results given above.

For each fermion $\psi_j(x)$ we have two $2\pi$-periodic bosonic fields $\varphi_j(x)$ and $\theta_j(x)$. They satisfy $[\partial \varphi_j(x),\theta_j(y)] = 2\pi\rmi \delta(x-y)$, and we have corresponding local vertex operators $V^{(j)}_{m,n} = \exp(\rmi m\varphi_j(x) +\rmi n \theta_j(x))$ for $m,n\in\mathbb{Z}$, these are primary fields with scaling dimension $h+\overline{h} = m^2 +n^2/4$ and conformal spin $h-\overline{h}=m n$. Fermionic operators are of the form $V^{(j)}_{m,n}$ with $m$ half-integer---this corresponds to a Jordan-Wigner string and arises from integrating the density fluctuation $\partial \varphi_j(x)/2\pi$. More general vertex operators can be written as products of the above for different sectors, and the scaling dimension is then a sum over the scaling dimension from each sector. 

Now, when $\alpha = \pi$, the operator $\rme^{\rmi \alpha Q_A}=\rme^{\rmi \alpha \langle Q_A\rangle} \rme^{\rmi \alpha Q_\Delta} $ corresponds to the Jordan-Wigner string. Similarly, for general $\alpha$ we can write down a corresponding operator: 
\begin{align}
\rme^{\rmi \alpha Q_\Delta} = \exp\left( \sum_{j=1}^N\left(\frac{\rmi\alpha}{2\pi}\int_{0}^{L} \partial\varphi_j(x) ~\rmd x \right)\right).\end{align}
 If we take the vacuum expectation, this looks like a two point function of vertex operators $\mathcal{V}_\alpha(x)= \exp\left( \sum_{j=1}^N\frac{\rmi\alpha}{2\pi}\varphi_j(x)\right)$ at $x=0$ and $x=L$. Using the usual formula for the scaling dimension, we have $h_\alpha = \overline{h}_\alpha = \frac{N}{2} \left(\frac{\alpha}{2\pi}\right)^2$. These operators are, of course, not local fields in the bosonic CFT; these nonlocal operators generically map between different sectors with different boundary conditions \cite{Ginsparg90,Alcaraz92}. Indeed, focusing on the $c=1$ complex fermion CFT, the operator:
\begin{align}
\mathcal{V}_\alpha(z) = \rme^{\rmi \frac{\alpha}{2\pi}( \varphi(z) - \varphi(0))} =\rme^{\rmi \frac{\alpha}{2\pi} \int_0^z \partial \varphi(z) \rmd z } 
\end{align} 
corresponds to a cut (along the integration contour) where the fermionic field (on the plane) has a phase shift of $\rme^{-\rmi \alpha}$. The scaling dimension of this operator, for $-\pi<\alpha<\pi$, then corresponds to the ground state energy of the fermionic CFT with corresponding twisted boundary condition $\psi(x+2\pi) = -\rme^{-\rmi \alpha} \psi(x)$ on a cylinder of radius one \cite{Ginsparg90,Polchinski98}. A thorough treatment of these twisted boundary conditions of the CFT is given in \cite[Chapter 10]{Polchinski98}---the key result is that we have $h_\alpha =\overline{h}_\alpha = \frac{1}{2}\left(\frac{\alpha}{2\pi}\right)^2$; and in fact this holds for $\alpha \in \mathbb{R}$.  Note also that as we take $\alpha \rightarrow \pm \pi$ we have degenerate CFT ground states: this is the usual degeneracy in the scaling dimensions of $\rme^{\pm \rmi \varphi/2}$, and means that we should expect that the dominant asymptotics changes at these values of $\alpha$. 

Returning to the $c=N$ CFT, the field $\mathcal{V}_\alpha(x)= \exp\left( \rmi\sum_{j=1}^N\frac{\alpha}{2\pi}\varphi_j(x)\right)$ gives the same twisted boundary condition for each of the different fermions, and the scaling dimension is, as expected, $h_\alpha =\overline{h}_\alpha = \frac{N}{2}\left(\frac{\alpha}{2\pi}\right)^2$. Note that the more general operator
$\mathcal{V}_{\alpha;n_1,\dots,n_N}(x)= \exp\left(\rmi \sum_{j=1}^N\left(\frac{\alpha}{2\pi}+n_j\right)\varphi_j(x)\right)$ creates the same boundary conditions; but for $-\pi <\alpha<\pi$ corresponds to an excited state in that sector (i.e., it is subdominant to $\mathcal{V}_\alpha(x)$) since it has scaling dimension $h_\alpha =\overline{h}_\alpha = \sum_{j=1}^N\frac{1}{2}\left(\frac{\alpha}{2\pi}+n_j\right)^2$.
 \subsubsection{The formula for the charged moment}
 
Using \eqref{eq:CFTformula}, with the above choice of operator $\mathcal{V}_\alpha$, leads to:
 \begin{align}
Z_n(\alpha) = \textrm{const}\times   L_{0}^{-\frac{N}{6}(n-1/n) -\frac{2N}{n} \left(\frac{\alpha}{2\pi}\right)^2}\rme^{\rmi \alpha \langle Q_A\rangle }(1+o(1)), \label{eq:CFTmoment}
\end{align}
where $L_{0}=L/a_0$ for some constant $a_0$ with dimensions of length. The other constant corresponds to the normalisation of the twist field. 
\subsubsection{Subdominant terms}
In fact, the composite twist operator will generically contain all terms allowed by symmetry. This leads to an expansion of the form:
\begin{align}
\mathcal{T}_{n,\alpha}=&\mathcal{T}_{n}(x) \times \Bigg (\sum_{\{n_j\}\in \mathbb{Z}^N} \Big(c_{\{n_j\},x}\rme^{\rmi \sum_{j=1}^N \left(\frac{\alpha}{2\pi}+n_j\right)\varphi_j(x) } +\mathrm{descendants~of~this~operator}\Big)\nonumber\\&+\sum_{\{n_j\} \in \mathbb{Z}^N}\sum_{\{m_j\} \in S}\Big(c_{\{m_j,n_j\},x} \rme^{\rmi \sum_{j=1}^N \Big(\left(\frac{\alpha}{2\pi}+n_j+\frac{m_j}{2}\right)\varphi_j(x)+ m_j\theta_j(x)\Big)}  +\mathrm{descendants~of~this~operator}\Big)\Bigg),  \label{eq:descendants}
\end{align}
where the constants $c_{\{n_j\},x}$ contain the relevant normalisations and also oscillatory terms (these can be obtained by replacing $\varphi_j(x) \rightarrow \varphi_j(x)+k_j x$ or $\varphi_j(x) \rightarrow \varphi_j(x)-k_j x$ as appropriate). The set $S$ corresponds to $m_j \in \mathbb{Z}$, such that at least two of the $m_j$ are not equal to zero, and $\sum_{j=1}^N m_j = 0$.
Note that this second sum does not occur in the $N=1$ case; it corresponds to overall uncharged terms that create fermions in one sector and annihilate fermions in another. Note also the periodicity in $\alpha$ in the set of fields that we know holds for $Z_n(\alpha)$.

 Inserting this ansatz into \eqref{eq:CFTformula}, we find two possibilities for the order of the dominant subleading term:
 \begin{enumerate}
 \item The first comes from subdominant primary operators with $o(1)=O\left(L_{0}^{-2d(n,\alpha)}\right)$ where $d(n,\alpha) =\frac{1}{n}\left(1-\frac{\lvert \alpha\rvert}{\pi} \right)$. There are a number of operators with this scaling dimension. First, we have contributions from the first sum in \eqref{eq:descendants} where one of the $n_j = -\mathrm{sign}(\alpha)$ and the rest are zero. For $N>1$, we also have contributions at the same order from the second sum. For $j_1\neq j_2$ put $m_{j_1}=1$, $m_{j_2}=-1$ and $n_{j_1}=-(\mathrm{sign}(\alpha)+1)/2$ and $n_{j_2}=-(\mathrm{sign}(\alpha)-1)/2 $, leading to the same scaling dimension. Note that more generally if our system is described by a compact boson away from the free-fermion radius one of these contributions will be subdominant compared to the other.
\item The second corresponds to contributions of descendants; the dominant descendant correction will satisfy $o(1)=O(L_{0}^{-1})$. 
 This is not immediate from \eqref{eq:descendants}; indeed the correspondence for the scaling dimensions \eqref{eq:scalingdim} derived in Ref.~\cite{Goldstein18} relies on the property that the field is primary, so we should not apply that formula. It would be interesting to derive the behaviour of transformed descendants either in the same way, or using the partition function approach \cite{Xavier18}; see also discussion in Ref.~\cite{Berganza2012}. We simply note that the $\mathbb{Z}_n$-orbifold CFT will have fields arranged into primaries and descendants, and so we expect the descendant fields to map onto the corresponding tower of states. This is consistent with the findings of \cite{Calabrese10}. \end{enumerate}
Using the formula \eqref{eq:mu}, we see that the CFT prediction for the subleading term is $O(L^{-2\mu(n,\alpha)})$, while our rigorous bound is $O(L^{-\mu(n,\alpha)})$---these are consistent. The CFT prediction for the subleading term matches numerical calculations in the XX model \cite{Bonsignori19}. Note that these numerical calculations were for $n=2$, and for $n\geq 2$ the contribution from the subleading vertex operator always dominates the contribution from the first descendant. We will discuss these subdominant contributions further in Section \ref{sec:FH}.

Finally, for the XX model case, we note that a formula was found for the expansion coefficients $c_{\{n_j\},x}$ in \eqref{eq:descendants} by an analytic continuation in $\alpha$ of the function $\Upsilon(n,\alpha)$ \cite{Fraenkel20}. This agrees with the approach based on Fisher-Hartwig expansions given below in \eqref{eq:subleading2}. 
\subsection{Beyond field theory: form factors and twisted overlaps}\label{sec:formfactor}
Estienne, Ikhlef and Morin-Duchesne analysed the problem of symmetry-resolved entropies in the XXZ model, which includes as a special case the XX model at half-filling: $f(\rme^{\rmi k}) = \cos(k)$ \cite{Estienne21}. Of particular interest to our discussion is that the constants appearing in \eqref{eq:CFTmoment} can be identified with limits of certain twisted overlaps in the lattice model that can be calculated separately. This uses details of the lattice model, but is a separate approach to the one taken in this paper. Note in particular that this method relies on the CFT description of the low-energy behaviour, and the replica approach fixes $n\in\mathbb{Z}_+$.

Suppose we have our system on a periodic ring; the starting point is that the two-point function of twist operators in the imaginary time direction is, by the state operator correspondence, proportional to the square of the absolute value of the twisted overlap $ \langle\psi_0\vert\psi_\alpha \rangle $, where $\vert\psi_\alpha \rangle$ is the ground state on the ring with $\alpha$-twisted boundary conditions. We can replace this by the limit of the ground state of a \emph{chain} with $M$ sites and with $\alpha$-twisted boundary conditions, denoted $\vert\psi_\alpha(M) \rangle$. More precisely, \begin{align}
Z_1(\alpha) =  \lvert \mathcal{A}_1(\alpha)\rvert^2 \rme^{\rmi \alpha \langle Q_A\rangle} L^{-2\left(\frac{\alpha}{2\pi}\right)^2}+\dots\end{align} where we have the form-factor \begin{align}
\lvert \mathcal{A}_1(\alpha)\rvert^2 = \lim_{M\rightarrow\infty} \Bigg(\Bigg(\frac{M}{2\pi}\Bigg)^{2\left(\frac{\alpha}{2\pi}\right)^2} \lvert\langle\psi_0(M) \vert\psi_\alpha(M) \rangle \rvert^2\Bigg).
\end{align} 
For the half-filled XX model, direct calculation of the twisted overlap gives \cite{Estienne21}:
\begin{align}
\lvert \mathcal{A}_1(\alpha)\rvert^2 = 2^{-2\left(\frac{\alpha}{2\pi}\right)^2} \rme^{\Upsilon(1,\alpha)}.
\end{align}
In this model we have $\sigma=2$, so using the CFT formula above we recover the leading term given in Result \ref{resultone} (as had already been found in \cite{Bonsignori19,Fraenkel20}). The method can be extended beyond half-filling, but one would need to give an asymptotic analysis of certain functions that appear\footnote{These take the form $\sum_y \log[\mathrm{sinc}\left(\frac{\pi}{M}(y+a)\right)]$ for particular ranges of $y$ that depend on the filling.}. For general filling, we hence expect that the outcome of this analysis would give:
\begin{align}
\lvert \mathcal{A}_1(\alpha)\rvert^2 = \sigma^{-2\left(\frac{\alpha}{2\pi}\right)^2} \rme^{\Upsilon(1,\alpha)} =\Big(2\sin(k_F)\Big)^{-2\left(\frac{\alpha}{2\pi}\right)^2}\rme^{\Upsilon(1,\alpha)},
\end{align}
although a direct derivation of this relation is beyond our scope.

There is a related formula for $\lvert \mathcal{A}_n(\alpha)\rvert^2$, where we take the lattice equivalent of the $\mathbb{Z}_n$-orbifold construction. In particular, we take $n$ replicas of the quantum chain, and introduce twisted boundary conditions that cyclically connect the $j$th copy to the $(j+1)$th and implement the $\alpha$-twist. We take the overlap $\mathcal{A}_n(\alpha)$ of this twisted ground state and the ground state of $n$ decoupled chains, and then we have the following formula for the half-filled XX model \cite{Estienne21}:
\begin{align}
\lvert \mathcal{A}_n(\alpha)\rvert^2 &= 2^{-\frac{1}{6}(n-1/n) - \frac{2}{n}\left(\frac{\alpha}{2\pi}\right)^2} \rme^{\Upsilon(n,\alpha)}, \end{align}
where the form \eqref{eq:upsilon} for $\Upsilon(n,\alpha)$ appears directly from the asymptotic analysis. Then:
\begin{align} Z_n(\alpha) &=  \lvert \mathcal{A}_n(\alpha)\rvert^2 \rme^{\rmi \alpha \langle Q_A\rangle} L^{-\frac{1}{6}(n-1/n) -\frac{2}{n}\left(\frac{\alpha}{2\pi}\right)^2}+\dots.\end{align}
 Again, since $\sigma =2$ this agrees with Result \ref{resultone} in this case. More generally, comparing this analysis to Result \ref{resultone} gives a formula for the limit of the twisted overlap for a general model in our class.
\begin{samepage}
 \subsection{Connection to our results}\label{sec:connection}
 \subsubsection{CFT ground states and Fermi velocities}
We can now use the CFT analysis to put our results in context. First, note that the CFT description \emph{of the low-energy physics} holds only for the case that all $v_j$ are equal (otherwise there is no rescaling so that space and time are symmetric). However, as pointed out in Section \ref{sec:class}, everything that we calculate is a ground state property, and the ground state is invariant if we deform our Hamiltonian by multiplying $f(\rme^{\rmi k})$ by a function\footnote{We can write $h(k) = \sum_{n=0}^R b_n \cos(nk)$ where $R$ corresponds to a minimal hopping range in the corresponding model. Hence, if we restrict to a finite-range class of models, we should have $R<\infty$.} $h(k)=h(-k)>0$. This means that for a given model, we can find a family of models with different Fermi velocities and different low-energy physics that nevertheless have the same ground state. In Appendix \ref{app:tuning} we explore this for the case $N=2$. There we find that along with the class of Hamiltonians that clearly have a low-energy CFT description (models with $k_2 = \pi-k_1$ and $h(k)=1$), we can also find a Hamiltonian that is a CFT at low energies and that shares the ground state with the model \eqref{eq:standardform} for \emph{any} $\pi/3<k_1<k_2<2\pi/3$. In general, due to the fact we can tune the Fermi velocities, we should expect that they do not enter into ground state properties, and moreover that we should expect to apply the CFT analysis above in general (although it remains to be shown explicitly that every model in our class shares its ground state with  another Hamiltonian in our class that has a low-energy CFT description). This general point is made for Bethe Ansatz solvable models in Ref.~\cite{Izergin89}. \end{samepage}
\subsubsection{Result \ref{resultone}: the charged moment}
We now discuss Result \ref{resultone} in light of the CFT analysis. Note that given Result \ref{resultone}, Results \ref{resulttwo} and \ref{resultthree} follow as in the lattice calculations by taking the Fourier transform. Comparing \eqref{eq:CFTmoment} to Result \ref{resultone}, we see that we have rigorous lattice asymptotics that agree with the scaling dimension predicted by the branch-point twist field analysis. We can identify the UV scale as $a_0 = a/\sigma$, where $a$ is the lattice spacing, and the composite twist operator $\mathcal{T}_{n,\alpha}$ has a normalised two-point function if we multiply by $\rme^{-N\Upsilon(n,\alpha)/2}$. As expected, both the lattice results and the form of the CFT results are independent of the Fermi velocities $v_j$. We have already compared the error term in Result \ref{resultone} to the error term predicted by CFT; we will look at this more deeply in Section \ref{sec:FH} where we discuss corrections to the asymptotics of the Toeplitz determinant. 

It is interesting that, aside from the mean charge, the dependence on the Fermi momenta, $k_j$, enters into the length scale $a_0$ only. A similar point was observed for the full counting statistics, $Z_1(\alpha)$, in Ref. \cite{Abanov11}. We can interpret this as follows, fixing $N=1$, then, for a given $n$ and $\alpha$, the normalisation of the composite twist operator is universal. For $N>1$, we have a natural decoupling of the ground state into $N$ sectors, and the normalisation of the composite twist operator is again universal and corresponds to the $N=1$ normalisation in each low energy sector. Indeed, if we assume universality then this must be the case. This is because there is a lattice limit of decoupled chains where this decoupling is manifest through the entire calculation, and so this product normalisation appears in that case. We discuss this limit further in Section \ref{sec:decoupled}. 
\subsubsection{Reconstructing Result \ref{resultone}}
We now show how we can reconstruct the leading term of Result  \ref{resultone} using CFT analysis and previous lattice results, if one makes certain physically reasonable assumptions. In particular, let us suppose that normalisation of the twist operator for fixed $n$ and $\alpha$ is universal and that there is a single UV scale $a_0$ that depends only on the Fermi momenta characterising the ground state. Then we can state Result \ref{resultone} (for integer $n$ and with the CFT estimate for the error term) using $\Upsilon(n,\alpha)$ and $a_0=a/\sigma$. An analytic continuation would then recover the formula for all $n$. $\Upsilon(n,\alpha)$ is known from work on the XX model: for general $n$ using Toeplitz determinant methods \cite{Bonsignori19, Fraenkel20}, alternatively, for positive integer $n$, using the twisted overlap method \cite{Estienne21}. The scale $a_0$ was implicit in the result for the asymptotic von Neumann entanglement entropy given by Keating and Mezzadri \cite{Keating2004}, again using Toeplitz determinant methods. Note that the proof of Result \ref{resultone} does not use any of these assumptions, and therefore confirms the above reasoning. We also see in the (conjectured) structure of the subleading corrections given in Appendix \ref{app:phases} that the same universality properties do not appear to hold in general. In particular, unlike the normalisation of the leading term, the coefficients appearing in the operator expansion depend on the Fermi momenta.
\subsection{Decoupled chains}\label{sec:decoupled}
It is instructive to consider the limit of fully decoupled chains---these are models of the form \begin{align}
f(\rme^{\rmi k}) = \cos(N k) - h 
\end{align}  and are critical for $\lvert h \rvert <1$. [Note that for even values of $N$ these models are not in the class explicitly analysed above, since $f(\rme^{\rmi \pi})>0$. Recall that \eqref{eq:-f(z)} allows us to proceed in this case.] This gives $N$ decoupled copies of the XX model\footnote{To be precise, the model with Hamiltonian equivalent to $f(\rme^{\rmi k}) =  \cos(k) -h$.}, each with $k_F= \arccos(h)$.

We can compare the R\'enyi entropies from these two points of view. First, evaluating the entropies in our class using Result \ref{resultone} we have the asymptotic formula:
\begin{align}
S_n =  \Bigg(\frac{N}{6}\left(1+\frac{1}{n}\right)\log( L_\sigma)  + \frac{N}{1-n} \Upsilon (n) \Bigg)+o(1).
\end{align}
Then, using the results for the XX model \cite{Jin2004}, we have:
\begin{align}
S_n &= \sum_{r=1}^N \Bigg(\frac{1}{6}\left( 1+\frac{1}{n}\right)\log(2\sin(k_F) L/N) +\frac{1}{1-n}\Upsilon(n) \Bigg)+o(1)\nonumber\\
&=\frac{N}{6}\left( 1+\frac{1}{n}\right)\log(2\sin(k_F) L/N) +\frac{N}{1-n}\Upsilon(n)+o(1).
\end{align}

Note the universal normalisation of the composite twist operator appearing in both formulae. Comparing the two formulae, we have an identity between the length scales:
\begin{align}
 \Bigg(\prod_{r=1}^N 2 \sin(k_r) \prod_{1\leq r<s\leq N}  
\Bigg(\frac{\sin^2 (\frac{k_r+k_s}{2})}{\sin^2 (\frac{k_r-k_s}{2})} \Bigg)^{(-1)^{r+s}}\Bigg)^{1/N}=\frac{2\sin(k_F)}{N}, \label{eq:identity}
\end{align}
where $k_r$ are the solutions of $\cos(Nk)= h$ between 0 and $\pi$. 

We point out that this identity can be proved independently. Using the explicit form of solutions of $\cos(Nk)= \cos(\alpha)$:
\begin{align}
k_{2n+1} = \frac{\alpha}{N} +\frac{2\pi n}{N} \quad& n=0,\dots, \lceil N/2\rceil -1 \qquad \mathrm{and} \qquad
k_{2n} = - \frac{\alpha}{N} +\frac{2\pi n}{N} \quad& n=1,\dots, \lfloor N/2\rfloor,
\end{align}
one can show that the identity \eqref{eq:identity} follows from the simpler identity
\begin{align}
\sin \big({\alpha}/{N}\big)\prod_{s=1}^{N-1}\Bigg(\frac{
{\sin \big(\frac{\alpha}{N}+ \frac{\pi s}{N}\big)}}{{\sin \big(\frac{\pi s}{N}\big)}}\Bigg)= \frac{\sin(\alpha)}{N}.\end{align}
This in turn follows from $\sin(N z) = 2^{N-1} \prod_{k=0}^{N-1} \sin(z+\frac{k \pi}{N})$ \cite[Eq.~4.21.35]{NIST:DLMF}. 
\subsection{Discussion of Fisher-Hartwig determinants, generalisations and the error term for the charged moment}\label{sec:FH}
\subsubsection{The conjectured full asymptotic expansion}
The theory of Toeplitz determinants with Fisher-Hartwig symbols gives the leading term in the asymptotics of the determinant where the symbol has certain singularities; this includes the jump discontinuities that appear at the zeros of $f(z)$ above. The formula for this leading term was the long-standing Fisher-Hartwig conjecture \cite{Fisher1969,Hartwig1969}; this conjecture and further generalisations, have now been proved---see the comprehensive review \cite{Deift2013} and references therein. Notice that the canonical form \eqref{eq:canonical} is invariant if we put $\beta_r \rightarrow \beta_r+n_r$ such that $\sum_{r=1}^{2N}n_r =0$. Above, we chose the unique canonical form, or representation, with $\lvert\lvert\beta\rvert\rvert <1$; this gives the correct asymptotics, as proved by Basor \cite{Basor79}. In the case that $\lvert\lvert\beta\rvert\rvert =1$, Basor and Tracy proposed the generalised Fisher-Hartwig conjecture (gFHC) \cite{Basor1991}: this states that the leading asymptotics are a sum over the terms corresponding to each canonical form with $\lvert\lvert\beta\rvert\rvert =1$ . This generalised conjecture was proved in Ref.~\cite{Deift2011}. Note that for a symbol with Fisher-Hartwig singularities, there always exists a canonical form with  $\lvert\lvert\beta\rvert\rvert \leq 1$, and it is unique when $\lvert\lvert\beta\rvert\rvert < 1$ \cite{Deift2011}.

These results give the first term in the asymptotic expansion. In our analysis above we used the rigorous results of \cite{Deift2014} to give an estimate of the correction term. In the analysis of symmetry-resolved entanglement entropies in the XX model, given in Refs.~\cite{Bonsignori19,Fraenkel20}, conjectured forms of the sub-leading terms in the expansion are used in the analysis (although as we will see below, the dominant sub-leading terms are in fact a result of Kozlowski \cite{Kozlowski08}). The conjectural correction terms come from a full asymptotic expansion---this is also referred to as the gFHC, although it is distinct from the gFHC of Basor and Tracy mentioned above. For the discussion here it suffices to consider the symbol of Section \ref{sec:asymptotics}, fixing $N=1$ (this is the XX model case): 
\begin{align}
\rme^{V(z,\lambda)}&= \rme^{V_0(\lambda)} =  (\lambda+1) \left(\frac{\lambda+1}{\lambda-1}\right)^{-k_F/\pi}\\
\beta_2 &= -\beta_1 = \beta(\lambda)  =  \frac{1}{2\pi\rmi} \log\left(\frac{\lambda+1}{\lambda-1}\right); \qquad\qquad -\pi \leq \arg\left(\frac{\lambda+1}{\lambda-1} \right) <\pi.
\end{align}
This representation has  $\lvert\lvert\beta\rvert\rvert < 1$, and we can find other representations by taking $\beta(\lambda) \rightarrow \beta(\lambda)+m$ for $m\in\mathbb{Z}$.
Then the conjectured full asymptotic expansion of $D_L[t(z,\lambda)]$, for this case, is as follows:
\begin{align}
\underbrace{\sum_{m\in\mathbb{Z}} \Bigg(L_\sigma^{-2(\beta(\lambda)+m)^2} \rme^{(V_0-2\rmi k_F m) L } G(1+\beta(\lambda)-m)G(1-\beta(\lambda)+m)\overbrace{\left(1 + \sum_{k=1}^\infty a^{(m)}_k(\lambda) L_\sigma^{-k}\right)}^{\textrm{`Descendant' corrections~for~each~}m}\Bigg)}_{\textrm{Sum~over~Fisher-Hartwig~representations}},\label{eq:fullexpansion}
\end{align}
where $L_\sigma=2\sin(k_F)L $ and the coefficients $a^{(m)}_k(\lambda)$ are not specified. Using the idea of a Fisher-Hartwig representation \cite{Bottcher2006,Deift2011}, one can write down a general conjecture of this form for \emph{any Toeplitz determinant with Fisher-Hartwig singularities}. This expansion is natural from the perspective of conformal field theory. For example, in the study of spin-correlations in the XX model, Toeplitz determinants with Fisher-Hartwig singularities appear (see \cite{Jones2019} and references therein); then the sum over Fisher-Hartwig representations corresponds to a sum over vertex operators \cite{Affleck88,Ginsparg90,Fisher97,Polchinski98,diFrancesco99,Sachdev01} and the correction terms correspond to the descendants. Another example is the sum over vertex operators in the analysis above---see Eq.~\eqref{eq:descendants}---however, the comparison is less direct since we must analyse the contour integral as in Section \ref{sec:proofone} to go from one expansion to the other. 
In Appendix \ref{app:phases} we compare the dominant subleading terms using the two approaches for the case $N=2$. We see that they agree both in the power of $L$ that appears, as well as in the oscillatory factors.
There are many examples in the literature of using subdominant terms that arise from these Fisher-Hartwig representations with $\lvert\lvert\beta\rvert\rvert >1$, for example Refs.~\cite{Abanov03,Franchini05,Calabrese10,Abanov11,Gutman11,Ivanov13b}. General conjectures of the form \eqref{eq:fullexpansion} are found in Refs.~\cite{Kozlowski08,Kitanine09,Calabrese10,Ivanov13,Ivanov13b,Stephan14}. Note that in some cases the term gFHC refers to the sum without the descendant corrections, while in other places it refers to the whole expansion. Indeed, in a study of corrections to scaling in entanglement entropies  \cite{Calabrese10}, the the term gFHC is used to refer to the sum over representations without the descendant correction terms, but the authors expect and indeed derive such correction terms---we will use their result for the first descendant correction below. 
\subsubsection{Kozlowski's results for the subleading term}
Now, while the full expansion is conjectural, we emphasise that Ref.~\cite{Kozlowski08} gives a derivation of the first subleading corrections for symbols with $\lvert\lvert \beta\rvert\rvert<1$ using Riemann-Hilbert techniques. This is most interesting and indeed agrees with the more general conjecture given above. Moreover, these first subleading terms are already enough to make a non-trivial comparison to field theory predictions, and naturally coincide with the subleading terms analysed in the recent works \cite{Bonsignori19,Fraenkel20} on symmetry-resolved entanglement entropy in the XX model. 

More precisely, in Ref.~\cite{Kozlowski08}, the first subleading terms in the asymptotic expansion of $D_L[t(z,\lambda)]$ were found. These include both oscillatory terms and non-oscillatory terms; and the oscillatory terms can be identified as corresponding to subdominant Fisher-Hartwig representations. Indeed, this motivated Kozlowski to give a general conjecture for an expansion of the form \eqref{eq:fullexpansion}. We also mention that a closely related formula, at least for the oscillatory terms, is found in \cite{Deift2011}. The result in Ref.~\cite{Kozlowski08} is an expansion of the form:
\begin{align}
\log D_L[t(z,\lambda)] = V_0(\lambda) L - 2N\beta(\lambda)^2 \log L + \log E(\lambda)+ \frac{A(L,\lambda)}{L^2} (1+o(1)) + \frac{B(\lambda)}{L} (1+o(1)),\label{eq:subleading}
\end{align}
where $A(L,\lambda)$ is the oscillatory correction (the $L$ dependence is purely oscillatory) and $B(\lambda)$ is the non-oscillatory correction.
Specialising the general formulae to our case (in particular, setting up the problem as in Result \ref{resultone}) the oscillating corrections are given by:
\begin{align}
A(L,\lambda) =& \sum_{1\leq r <s\leq 2N}\Bigg(\frac{\Gamma(1-\beta_r) \Gamma(1+\beta_s)}{\Gamma(\beta_r)\Gamma(-\beta_s)}L^{2\beta_r -2\beta_s } \left(\frac{z_s}{z_r}\right)^L \frac{1}{\lvert z_s-z_r\rvert^2}\frac{\prod_{t\neq r} \lvert z_t-z_r\rvert^{-2\beta_t}}{\prod_{t\neq s}\lvert z_t-z_s\rvert^{-2\beta_t}}\nonumber\\
 &+\frac{\Gamma(1-\beta_s) \Gamma(1+\beta_r)}{\Gamma(\beta_s)\Gamma(-\beta_r)}L^{2\beta_s -2\beta_r } \left(\frac{z_r}{z_s}\right)^L \frac{1}{\lvert z_s-z_r\rvert^2}\frac{\prod_{t\neq s} \lvert z_t-z_s\rvert^{-2\beta_t}}{\prod_{t\neq r}\lvert z_t-z_r\rvert^{-2\beta_t}}\Bigg),\label{eq:oscillating}
\end{align}
where the $\lambda$ dependence is through the $\beta_r$. Recall that the $z_k$ are defined in Section \ref{sec:asymptotics}. Note that we can group the sum depending on the power of $L$ appearing; there are three terms, $L^{4\beta(\lambda)}$, $L^0$ and $L^{-4\beta(\lambda)}$:
\small{
\begin{align}
&A(L,\lambda) = \sum_{r=-\lceil N/2\rceil+1}^{\lfloor N/2\rfloor}\sum_{s=-\lfloor N/2\rfloor}^{\lceil N/2 \rceil -1}\frac{\Gamma(1-\beta(\lambda))^2}{\Gamma(\beta(\lambda))^2}L^{ 4\beta(\lambda) } \left(\frac{z_{N+2r}}{z_{N+1+2s}}\right)^L \frac{1}{\lvert z_{N+2r}-z_{N+1+2s}\rvert^2}\frac{\prod_{t\neq N+1+2s} \lvert z_t-z_{N+1+2s}\rvert^{-2\beta_t}}{\prod_{t\neq N+2r}\lvert z_t-z_{N+2r}\rvert^{-2\beta_t}}\nonumber\\
 &+\sum_{r=-\lceil N/2\rceil+1}^{\lfloor N/2\rfloor}\sum_{s=-\lfloor N/2\rfloor}^{\lceil N/2 \rceil -1}\frac{\Gamma(1+\beta(\lambda))^2}{\Gamma(-\beta(\lambda))^2}L^{-4\beta(\lambda)} \left(\frac{z_{N+1+2s}}{z_{N+2r}}\right)^L \frac{1}{\lvert z_{N+2r}-z_{N+1+2s}\rvert^2}\frac{\prod_{t\neq N+2r} \lvert z_{t}-z_{N+2r}\rvert^{-2\beta_t}}{\prod_{t\neq N+ 1+2s}\lvert z_t-z_{N+1+2s}\rvert^{-2\beta_t}} +O(L^0). \label{eq:A} \end{align}}
 \normalsize
The $O(L^0)$ term is not needed for the analysis of the dominant subleading terms, but can be written down explicitly if needed using \eqref{eq:oscillating}. Furthermore, the non-oscillating corrections are given by:
\begin{align}
B(\lambda) = -\beta(\lambda)^2\sum_{r=1}^{2N}  \left(\sum_{s \neq r}\beta_s \frac{(z_r+z_s)}{(z_r-z_s)} \right) = \beta(\lambda)^3\sum_{r=1}^{2N}  \left(\sum_{s \neq r}(-1)^{N-s} \frac{(z_r+z_s)}{(z_r-z_s)} \right) . \label{eq:B}
\end{align}

In Ref.~\cite{Kozlowski08}, the expansion \eqref{eq:subleading} is derived with no control over the error terms. Let us suppose that \eqref{eq:subleading} holds with $o(1)$ terms that are uniform in $\lambda$. Then we can integrate to find explicit subleading corrections to the charged moment. In particular, with the usual definitions as in Result \ref{resultone}, we would have:
\begin{align}
&\log Z_n(\alpha,L) =\nonumber\\& \Bigg( -\frac{N}{6}\left(n-\frac{1}{n}\right)-\frac{2N}{n} \left(\frac{\alpha}{2\pi}\right)^2 \Bigg) \log( L_\sigma) +  {\rmi \alpha \langle Q_A\rangle}+{N \Upsilon(n,\alpha)}  +d_n(\alpha,L)(1+o(1))+\tilde{d}_n(\alpha,L)(1+o(1)) \end{align} 
where 
\begin{align}
d_n(\alpha,L) &=-\frac{1}{2\pi\rmi L^2} \int_{C_0(\eps,\delta)} \frac{\rmd f_n(1+\eps,\lambda,\alpha) }{\rmd \lambda} A(L,\lambda)\rmd \lambda\\
\tilde{d}_n(\alpha,L) &=-\frac{1}{2\pi\rmi L} \int_{C_0(\eps,\delta)} \frac{\rmd f_n(1+\eps,\lambda,\alpha) }{\rmd \lambda} B(\lambda) \rmd \lambda . \end{align} 
Defining \begin{align}
\sigma_\star(r,s) = \Bigg(\frac{\prod_{t\neq N+1+2s} \lvert z_t-z_{N+1+2s}\rvert^{(-1)^{N-t}}}{\prod_{t\neq N+2r}\lvert z_t-z_{N+2r}\rvert^{(-1)^{N-t} }}\Bigg)^{1/2} >0,
\end{align}
we show in Appendix \ref{app:subleading} that $d_n(\alpha,L)$ has the following form:
\begin{align}
&d_n(\alpha,L) = L^{-\frac{2}{n}(1+\alpha/\pi)}\frac{\Gamma(\frac{1}{2}+\frac{1}{2n}(1+\alpha/\pi))^2}{\Gamma(\frac{1}{2}-\frac{1}{2n}(1+\alpha/\pi))^2}\Bigg( \sum_{r=-\lceil N/2\rceil+1}^{\lfloor N/2\rfloor}\sum_{s=-\lfloor N/2\rfloor}^{\lceil N/2 \rceil -1}\left(\frac{z_{N+2r}}{z_{N+1+2s}}\right)^L\frac{\sigma_\star(r,s)^{2-\frac{2}{n}(1+\alpha/\pi)}}{\lvert z_{N+2r}-z_{N+1+2s}\rvert^2}\Bigg) \nonumber\\
 &+ L^{-\frac{2}{n}(1-\alpha/\pi)}\frac{\Gamma(\frac{1}{2}+\frac{1}{2n}(1-\alpha/\pi))^2}{\Gamma(\frac{1}{2}-\frac{1}{2n}(1-\alpha/\pi))^2}\Bigg(\sum_{r=-\lceil N/2\rceil+1}^{\lfloor N/2\rfloor}\sum_{s=-\lfloor N/2\rfloor}^{\lceil N/2 \rceil -1} \left(\frac{z_{N+1+2s}}{z_{N+2r}}\right)^L \frac{\sigma_\star(r,s)^{2-\frac{2}{n}(1-\alpha/\pi)}}{\lvert z_{N+2r}-z_{N+1+2s}\rvert^2}\Bigg) +O(L^{-2}).\label{eq:subleading2}\end{align}
 For $n\geq 2$ one of the two explicit terms will be the dominant subleading term. Otherwise, one might be interested in the exact form of $\tilde{d}_n(\alpha,L)$ which is $O(L^{-1})$---see the discussion in the next section. Note that specialising this formula to the $N=1$ case, we recover the results of Refs.\cite{Bonsignori19,Fraenkel20}.
 
 We emphasise that this conclusion requires assuming uniformity of the error term. It would be interesting to prove this using the Riemann-Hilbert method, taking the results of Deift, Its, Krasovsky \cite{Deift2011,Deift2014} and Kozlowski \cite{Kozlowski08} further. Indeed, in Ref.~\cite{Deift2011} it is remarked that given an analytic $V(z)$ (in our case of interest $V(z)$ is constant), one can derive subleading terms in the expansion using the methods of that paper, i.e., using Riemann-Hilbert methods. It would be most interesting to calculate the Fisher-Hartwig subleading term this way, with control of the error term, and hence to rigorously derive the subleading terms for the charged moment in agreement with the CFT description.
\subsubsection{Descendant corrections}\label{sec:descendantcorrections}
The expression \eqref{eq:subleading2} gives the expected subdominant behaviour for $n\geq 2$ but there is a leading $1/L$ correction for $0< n< 2$ (for a range of $\alpha$); although the coefficient in the expansion could be zero. Note that the corrections for $N=1$ have been analysed in Ref.~\cite{Calabrese10} as follows. Reinterpreting the Toeplitz determinant of interest as a random matrix average, a recursion relation, determined by the Painlev\'e VI equation, for $D_L[t(z),\lambda)]$ at different values of $L$ can be used to derive subdominant terms \cite{Keating2004,Calabrese10}. Inserting the above ansatz \eqref{eq:fullexpansion} into the recursion relation, Calabrese and Essler found that $a^{(0)}_1=4\rmi\cos(k_F) \beta(\lambda)^3$. 
This moreover agrees with the result for non-oscillating corrections given by Kozlowski (in this case $B(\lambda)/L= 2\rmi \beta(\lambda)^3 \cot(k_F)/L =a^{(0)}_1/L_\sigma$). Using Eq.~\eqref{eq:contourint} and deforming to $C(\eps,\delta')$, the corresponding term of order $L_\sigma^{-1}$ is:
\begin{align}
\Big[\tilde{d}_n(\alpha,L)\Big]_{N=1}=&-\frac{2\cos(k_F)}{\pi  L_\sigma} \lim_{\eps,\delta'\rightarrow 0} \int_{-1}^1 \frac{\rmd}{\rmd\lambda}\Big( f_n(1+\eps,\lambda,\alpha)\Big)  \Big(\beta(\lambda+\rmi \delta)^3-\beta(\lambda-\rmi \delta)^3\Big)\rmd\lambda\nonumber
\\
&=-\frac{2\cos(k_F)}{  L_\sigma} \underbrace{n\int_{-\infty}^\infty \left(\tanh(\pi w)-\tanh(\pi nw+\rmi\alpha/2) \right) (3w^2-1/4) \rmd w}_{J(n,\alpha)},\label{eq:descendantintegral}
\end{align}
where in the second line we used standard substitutions as in Refs.~\cite{Calabrese10,Bonsignori19}. Note that $J(n,0)$ is the integral of an odd function of $w$ and thus vanishes. This is consistent with the results of Calabrese and Essler: the $O(L^{-1})$ correction is not present in their results for the R\'enyi entropies. (They do find descendant corrections of order $L^{-2}$, but these are always subdominant to the leading correction coming from a primary.)
However, in the charged case this integral is non-zero in general. Indeed in Appendix \ref{app:descendant} we prove that:
\begin{align}
J(n,\alpha) = \frac{\rmi}{4n^2} \left((n^2-1)\frac{\alpha}{\pi}  +\Big(\frac{\alpha}{\pi} \Big)^3 \right); \label{eq:J}
\end{align}
this is zero only for $\alpha = 0$ (for general $n$) and for $\alpha = \pm\pi\sqrt{1-n^2} $ (when $n\leq 1$). When $n=1$, this is a correction to the full-counting statistics generating function, and then \eqref{eq:J} agrees with the result found in \cite{Ivanov13b}. 

More generally, let us suppose that \eqref{eq:subleading} holds uniformly. Then we can insert \eqref{eq:B} into the contour integral and using \eqref{eq:J} we have that:
\begin{align}
\tilde{d}_n(\alpha,L)= -\frac{1}{8 n^2 L} \left((n^2-1)\frac{\alpha}{\pi}  +\Big(\frac{\alpha}{\pi} \Big)^3 \right)  \Bigg( \sum_{r=1}^{2N} \sum_{s\neq r} (-1)^{N-s} \frac{z_r +z_s}{z_r-z_s}  \Bigg).\label{eq:tilded}
\end{align}

 To conclude this discussion, we have seen a close relationship between the CFT expansion in primary operators and descendants and the asymptotic expansion of Toeplitz determinants with Fisher-Hartwig symbols, the latter based on subdominant Fisher-Hartwig representations. We emphasise that both of these techniques are in excellent agreement with each other, and indeed with numerical calculations in the XX model \cite{Calabrese10,Bonsignori19,Fraenkel20}. Moreover the form of the dominant subleading terms in the Fisher-Hartwig expansion is found in \cite{Kozlowski08}. However, the uniformity of the error term in that case has not been established, and so we have used weaker bounds on the error term in deriving our results above---this is enough to establish the expansion in powers of $\log(L)^{-1}$ for the symmetry-resolved entropies. It is an exciting question to understand more deeply these subdominant terms, and the relationship between the structure of the Fisher-Hartwig expansion and conformal field theory.
\section{Outlook}
We have obtained exact formulas for symmetry-resolved R\'enyi entropies in a large class of critical free-fermion lattice models. These lattice models have a low-energy description given by $N$ massless complex fermions. The key step was an asymptotic analysis of the charged moments using Toeplitz determinant theory, including a rigorous error term. We emphasise an effective scale $\sigma$, that depends only on the Fermi momenta. We have given the physical context for our results, and made connections between the CFT analysis and the theory of Toeplitz determinants with Fisher-Hartwig singularities. Our work significantly expands upon the work on the XX model given in Refs.~\cite{Bonsignori19,Fraenkel20}. We also give connections to the lattice methods based on twisted overlaps, given in Ref.~\cite{Estienne21}.

As far as open questions are concerned, a key issue is a greater understanding of subleading terms in the calculation of the charged moment. This is discussed at length in the previous section, we reemphasise here that the Riemann-Hilbert method, as used in Ref.~\cite{Kozlowski08,Deift2011,Deift2014}, could be used to rigorously derive these terms. 
It may be useful to separate the problem of proving the leading subdominant terms with uniform error estimates from the task of establishing the full asymptotic expansion.
Note also that the class we consider here, with $N>1$, involves an expansion in CFT fields that goes beyond that appearing in the $N=1$ case. We have presented calculations in Appendix \ref{app:phases} that show that in the case $N=2$, there is a correspondence between leading subdominant vertex operators and the leading subdominant Fisher-Hartwig representations. It would be most interesting to see if this were true in general.

The twisted overlaps used in Ref.~\cite{Estienne21} to derive the full lattice result from the CFT formula are themselves determinants of Toeplitz matrices. It would be interesting to derive the asymptotics of these overlaps for the more general class considered here directly; our results imply what these asymptotics should be. We also note that the form-factor calculations apply also to subdominant terms in the CFT expansion. It would be very interesting to see if they agree with the constants one would get from using subdominant Fisher-Hartwig representations.

The charged moment $Z_n(\alpha)$ is periodic in $\alpha$. The dominant asymptotic term is valid only when fixing the range $-\pi<\alpha<\pi$, which suffices for many purposes. Including the subleading corrections in the CFT expansion \eqref{eq:descendants} can in principle restore periodicity in $\alpha$---see relevant discussion in Refs.~\cite{Belin13,Fraenkel20,Estienne21}. Indeed in Ref.~\cite{Fraenkel20} an analytic (rather than periodic) continuation of $\Upsilon(n,\alpha)$ 
to $\lvert \alpha \rvert >\pi$ is found. This is again relevant to the subleading terms and both the CFT and Fisher-Hartwig expansions. We saw agreement with the dominant subleading primary terms in the XX model case. A similar comparison to the conjectured coefficients in the Fisher-Hartwig expansion should be made more generally. 

In this work we studied systems with U(1) symmetry. We can consider larger classes of models that contain our family, for example, the BDI class of topological superconductors \cite{Altland97,Ryu10,Verresen2018}. The gapless models considered here are at phase transitions of gapped phases that have symmetries including $\mathbb{Z}_2$ fermion parity symmetry. One can consider the corresponding $\mathbb{Z}_2$-resolved entropies; these were studied in \cite{Fraenkel20} for the XY spin chain. The entanglement entropies in the XY model can be analysed using Toeplitz determinant theory \cite{Its2007,Franchini07}. A more general analysis of all gapped models in the (translation-invariant) BDI class (equivalently, the Jordan-Wigner dual spin-1/2 chains) was given by Its, Mezzadri and Mo \cite{Its2008}. The results for the corresponding (block)-Toeplitz determinant are less simple in this case, but it would be interesting to understand if one could find the $\mathbb{Z}_2$-resolved entropies in that case. We also mention recent work on symmetry-resolved entropies in WZW models via CFT methods. The models considered here are simple examples of WZW models [the group is $\mathrm{SO}(2N)_1$]. One could try to find the corresponding symmetry-resolved entropies using the lattice model, confirming they fit the CFT prediction. 

We note that knowledge of the R\'enyi entropies for all $n$ is equivalent to knowing the entanglement spectrum \cite{Li08,Calabrese08,Pollmann10,Franchini10,Susstrunk12} (or spectrum of the reduced density matrix). It would be interesting to use our results to find the entanglement spectrum of the different symmetry sectors. 

Finally, to apply the CFT analysis to our class of models, we supposed that every ground state had a parent Hamiltonian that is described by a CFT at low energies. We proved this for a simple subclass of models with $N=2$, but this should apply in general based on the form of our results and the principles of \cite{Izergin89}. It would be neat to have a general formula for the function $h(k)$ that would allow us to constructively show this in the general case.

\section*{Acknowledgements}
I am grateful to Paul Fendley, Jon Keating, Karol Kozlowski and Ruben Verresen for illuminating discussions and correspondence. 

\bibliography{arxiv.bbl}

\appendix
\section{Details for Section \ref{sec:asymptotics}}\label{app:errorterm}
\subsection{The case \texorpdfstring{$n\geq2$}{n >2 and n=2}}\label{app:errorterm1}
Here we analyse $\lvert\lvert \beta\rvert\rvert$ on the contour of integration $C_0(\eps,\delta)$ (see Figure \ref{fig:contours}). Recall that $\lvert\lvert \beta\rvert\rvert$ is defined as
$\max_{j,k}  \lvert\Re(\beta_j) - \Re(\beta_k)\rvert$. Hence, for our choice of $\beta_j(\lambda)$ (given in \eqref{eq:beta}), we have that the seminorm $\lvert\lvert \beta(\lambda) \rvert\rvert$ is equal to $2\lvert\Re(\beta(\lambda))\rvert$.
Recall also that the branch points are at
\begin{align}
\rmi\lambda_+ &= \rmi (1+\eps)\tan\left(\frac{\pi-\alpha}{2n}\right)\\
\rmi\lambda_- &=-\rmi(1+\eps) \tan\left(\frac{\pi+\alpha}{2n}\right) .
\end{align}

Let us consider the vertical contours. First take $\lambda = (1+\eps/2)+ \rmi s$ for $(\lambda_- +\delta) \leq s \leq(\lambda_+-\delta)$; then:
\begin{align}
\lvert \Re\left(\beta(\lambda)\right)\rvert &= \frac{1}{2\pi}\Bigg\lvert \arctan\left(-\frac{2s}{s^2+\eps +\eps^2/4}\right)\Bigg\rvert ,
\end{align}
which is maximised at $s=\pm \sqrt{4\eps+\eps^2}$, with value $\frac{1}{2\pi}\lvert \arctan\left((\eps+\eps^2/4)^{-1/2}\right)\rvert$. The same formula holds for the other vertical contour, $\lambda = -(1+\eps/2)- \rmi s$. Hence, along the vertical contours we have that $\lvert \Re\left(\beta(\lambda)\right)\rvert \leq 1/4$, and so $\Big(\lvert\lvert \beta(\lambda) \rvert\rvert-1\Big) \leq -1/2$. 

The upper horizontal contour is $\lambda = \rmi (\lambda_+-\delta) -s$ for $-(1+\eps/2)<s <(1+\eps/2)$. We have:
\begin{align}
\Re(\beta(\lambda)) = \frac{1}{4\pi\rmi} \left(\log\left(\frac{\rmi (\lambda_+-\delta) -s+1}{\rmi (\lambda_+-\delta) -s-1} \right)-\log\left(\frac{\rmi (\lambda_+-\delta) +s-1}{\rmi (\lambda_+-\delta) +s+1} \right)\right).
\end{align}
It is straightforward to see (e.g., from \eqref{eq:betalambda} below) that $\Re(\beta(\lambda_+))<0$, and, moreover, that  $\Re(\beta(\lambda))$ has a minimum at $s=0$. Furthermore, $s=0$ is the only stationary point. Hence $\lvert\Re(\beta(\lambda))\rvert$
 is maximised at $s=0$, and so we have:
\begin{align}
\lvert \Re(\beta(\lambda_+))\rvert =\Bigg\lvert\frac{1}{2\pi\rmi} \log\left( -\left( \frac{1+\lambda_+}{1-\lambda_+} \right)\right) \Bigg\rvert= \Bigg\lvert\frac{1}{2\pi\rmi}\log\left( -\rme^{\rmi\theta_0}\right)\Bigg\rvert=\Bigg\lvert\frac{1}{2\pi} \left( \theta_0-\pi\right)\Bigg\rvert =\frac{1}{2\pi}\left(\frac{(n-1)\pi+\alpha}{n}\right).\label{eq:betalambda}
\end{align} 
Hence, along the upper horizontal contour we have:
\begin{align}
\lvert\lvert \beta(\lambda) \rvert\rvert -1 \leq  -\frac{1}{n} \left(1-\frac{\alpha}{\pi}\right) +O(\delta,\eps).\label{eq:uppercontour}
\end{align}

Now consider the lower horizontal contour with $\lambda = \rmi (\lambda_-+\delta) + s$ for $-(1+\eps/2)<s <(1+\eps/2)$. 
By analogous calculations, on the lower horizontal contour we have:
\begin{align}
\lvert\lvert \beta(\lambda) \rvert\rvert -1 \leq  -\frac{1}{n} \left(1+\frac{\alpha}{\pi}\right) +O(\delta,\eps).
\end{align}

Now, for $n\geq 2$ we have that  $-\frac{1}{n} \left(1-\frac{\lvert\alpha\rvert}{\pi}\right)\geq -1/2$; i.e., the horizontal contours dominate. We can then conclude that, for $-\pi<\alpha<\pi$ and along the contour $C_0(\eps,\delta)$, we have the following $\lambda$-independent bound:
\begin{align}
\lvert\lvert \beta \rvert\rvert -1 \leq -\frac{1}{n} \left(1-\frac{\lvert\alpha\rvert}{\pi}\right)  +O(\delta,\eps). \label{eq:betabound}
\end{align}
Note that for integer $n$ there is no branch cut on the real line, so we can extend the rectangular contour along the real axis. This will improve the bound on the vertical contour, but since this is always subdominant to the contribution from the horizontal contour it does not affect our result.

\subsection{The case \texorpdfstring{$0<n<2$}{0<n<2}}
We now have the possibility that there are no imaginary branch points either above or below the real axis, i.e., $\lambda_\pm = \pm \infty$. This means that we take our rectangular contour with corner at $\lambda =1+\eps/2 + \rmi R$ or  $\lambda =-1-\eps/2 - \rmi R$ respectively, and then take the limit of large $R$. Note that on the horizontal contour with imaginary part $\pm R$, we have $\Re(\beta(\lambda)) \rightarrow 0$. On the vertical contour, the maximum of $\Re(\beta(\lambda))$ is always at $s=\pm \sqrt{4\eps+\eps^2}$, independent of $R$. 

Depending on the value of $\alpha$ and $n$ we can use the above analysis to find a similar bound. First, let $0<n<1$ and $-\pi(1-n)<\alpha<\pi(1-n)$. Then $\rmi\lambda_\pm =\pm \rmi\infty$ and on the two horizontal contours we have $\Re(\beta(\lambda)) \rightarrow 0$. Thus from the vertical contours we have the  $\lambda$-independent bound:
\begin{align}
\lvert\lvert \beta \rvert\rvert -1 \leq -1/2.
\end{align}
For $0<n<1$ and $\pi(1-n)< \alpha<\pi$, we have a branch point at $\theta_0$, and close our contour below it. Then we have the bound $\eqref{eq:uppercontour}$ for the horizontal contour; and over the full contour we have the $\lambda$-independent bound 
\begin{align}
\lvert\lvert \beta \rvert\rvert -1 \leq \max\Bigg\{-\frac{1}{2},  -\frac{1}{n} \left(1-\frac{\lvert\alpha\rvert}{\pi}\right) \Bigg\} +O(\delta,\eps). \label{eq:betabound2}
\end{align}
The same bound follows for $0<n<1$ and $-\pi\leq \alpha\leq-\pi(1-n)$. Moreover, for $-\pi(1-n)<\alpha\leq\pi(1-n)$ we have $-\frac{1}{2}> - \frac{1}{n} \left(1-\frac{\lvert\alpha\rvert}{\pi}\right)$, so we conclude that \eqref{eq:betabound2} holds for $0<n<1$ and for $-\pi<\alpha<\pi$. 

By considering the different cases for $1< n<2$, we are again led to \eqref{eq:betabound2}. Hence, that bound holds for all $n>0$ and all $-\pi<\alpha<\pi$. For the special case $n=1$ we have a single branch point at $\theta_0$ for $0\leq \alpha <\pi$ and at $\theta_{-1}$ for $-\pi<\alpha<0$. There is no branch point on the real line, so we can extend the rectangular contour to $\pm\infty$ along the real axis. This means that on the vertical contours $\lvert \lvert \beta \rvert \rvert \rightarrow 0$, while the horizontal contours are analysed as above. We then see that for $n=1$:
\begin{align}
\lvert\lvert \beta \rvert\rvert -1 \leq \max\Bigg\{-1,  -\left(1-\frac{\lvert\alpha\rvert}{\pi}\right) \Bigg\} +O(\delta,\eps). \label{eq:betabound3}
\end{align}
In fact, for $\alpha=0$ this agrees with the expected CFT scaling, $O(L^{-1})$. However, based on the discussion in Section \ref{sec:FH}, we expect that this contribution has coefficient equal to zero (see Ref. \cite{Calabrese10} in the XX model case and \eqref{eq:tilded} more generally). Hence, once again, the rigorous bound does not give the expected (conjectural) leading subdominant term.
\section{CFT parent Hamiltonian for \texorpdfstring{$N=2$}{N=2}}\label{app:tuning}
Suppose we have a model of the form \eqref{eq:standardform}, with Fermi momenta $0<k_1<k_2<\pi$, and $h(k)=1$. The Fermi velocities are given by:
\begin{align}
v_1 &= \frac{1}{4} \sin(k_1) (\cos(k_1)-\cos(k_2))\nonumber\\
v_2 &=  \frac{1}{4} \sin(k_2) (\cos(k_1)-\cos(k_2)).
\end{align}
These are equal at all points $k_2 = \pi-k_1$.

Now, let $h(k) = 1+ b \cos(k)$, which is a valid choice of $h(k)$ for $\lvert b \rvert<1$; i.e., the Hamiltonian $f_1(\rme^{\rmi k})$ and the Hamiltonian $f_2(\rme^{\rmi k})=f_1(\rme^{\rmi k})h(k)$ have the same ground state. The Fermi velocities of the new model are those of the original model multiplied by $h(k)$ at the Fermi momenta, giving:
\begin{align}
v_1 &=\left(1+b\cos(k_1)\right) \frac{1}{4} \sin(k_1) (\cos(k_1)-\cos(k_2))\nonumber\\
v_2 &= \left(1+b\cos(k_2)\right) \frac{1}{4} \sin(k_2) (\cos(k_1)-\cos(k_2)).
\end{align}
We can then solve $v_1=v_2$ for $b$, giving:
\begin{align}
b= \frac{\sin(k_2)-\sin(k_1)}{\cos(k_1)\sin(k_1)-\cos(k_2)\sin(k_2)},
\end{align} 
which satisfies $\lvert b\rvert<1$ for $\pi/3<k_1<k_2 <2\pi/3$. 

Consider the analogous calculation using \begin{align}
h(k) = 1+\left(\frac{\sin(k_2)-\sin(k_1)}{\cos(nk_1)\sin(k_1)-\cos(nk_2)\sin(k_2)}\right)\cos(nk)
\end{align} 
for $n\geq 2$.
For this to be valid, we need 
\begin{align}
-1<\left(\frac{\sin(k_2)-\sin(k_1)}{\cos(nk_1)\sin(k_1)-\cos(nk_2)\sin(k_2)}\right)<1.
\end{align}
One can try to find a suitable $n$ satisfying this inequality for a given $k_1$ and $k_2$, or try different forms $h(k) = \sum_{n=0}^R b_n \cos(nk)$. A full analysis would be of interest since it would establish that for all models in our class, the ground states have a parent Hamiltonian that is a CFT at low energies. In this appendix we simply show that this conclusion holds beyond a measure-zero region of $k$-space.
\section{Details for Section \ref{sec:FH}}\label{app:subleading}
In this appendix we outline the steps to reach the explicit form of the subleading term given in Section \ref{sec:FH}. First, recall the definition:
\begin{align}
\sigma_\star(r,s) = \Bigg(\frac{\prod_{t\neq N+1+2s} \lvert z_t-z_{N+1+2s}\rvert^{(-1)^{N-t}}}{\prod_{t\neq N+2r}\lvert z_t-z_{N+2r}\rvert^{(-1)^{N-t} }}\Bigg)^{1/2} >0
\end{align}
and define corresponding length scale $L_\star = \sigma_\star L$. Then note that the expression for the oscillating corrections $A(L,\lambda)$ given in \eqref{eq:A} is a sum over terms of the form:
\begin{align}
J_1(r,s)=\frac{\rmi}{2\pi \lvert z_{N+2r}-z_{N+1+2s}\rvert^2 L^2} \left(\frac{z_{N+2r}}{z_{N+1+2s}}\right)^L \int_{C_0(\eps,\delta)} \frac{\rmd f_n(1+\eps,\lambda,\alpha) }{\rmd \lambda} \frac{\Gamma(1-\beta(\lambda))^2}{\Gamma(\beta(\lambda))^2} L_\star^{4\beta(\lambda)}\rmd \lambda \nonumber \\ 
J_2(r,s)= \frac{\rmi}{2\pi \lvert z_{N+2r}-z_{N+1+2s}\rvert^2L^2}  \left(\frac{z_{N+2r}}{z_{N+1+2s}}\right)^{-L}\int_{C_0(\eps,\delta)} \frac{\rmd f_n(1+\eps,\lambda,\alpha) }{\rmd \lambda} \frac{\Gamma(1+\beta(\lambda))^2}{\Gamma(-\beta(\lambda))^2} L_\star^{-4\beta(\lambda)} \rmd\lambda.
\end{align}
 To evaluate these integrals, deform the contour $C_0(\eps,\delta)$ to $C(\eps,\delta')$ and take the limit. We then reach standard integrals that have been studied before in the XX model case \cite{Calabrese10,Bonsignori19,Fraenkel20}. In particular:
\begin{align}
\lim_{\eps,\delta'\rightarrow 0} \int_{C(\eps,\delta')}& \frac{\rmd f_n(1+\eps,\lambda,\alpha) }{\rmd \lambda} \frac{\Gamma(1-\beta(\lambda))^2}{\Gamma(\beta(\lambda))^2} L_\star^{4\beta(\lambda)}\rmd \lambda\nonumber\\
&=\pi n \int_{-\infty}^{\infty} \left(\tanh(n\pi w+\rmi \alpha/2)-\tanh(\pi w) \right)\frac{\Gamma(\frac{1}{2}+\rmi w)^2}{\Gamma(\frac{1}{2}-\rmi w)^2} L_\star^{-4\rmi w+2} \rmd w \Big(1+O(L_\star^{-4})\Big)\nonumber\\
&= -2\pi\rmi L_\star^{2-\frac{2}{n}(1+\alpha/\pi)}\frac{\Gamma(\frac{1}{2}+\frac{1}{2n}(1+\alpha/\pi))^2}{\Gamma(\frac{1}{2}-\frac{1}{2n}(1+\alpha/\pi))^2} (1+o(1)).
\end{align} 
In the second line we use the change of variables $\lambda = \tanh(\pi w)$. The third equality follows from the residue theorem on closing the contour in the lower half-plane; the explicit leading term comes from the pole closest to the real axis.
Similarly
\begin{align}
\lim_{\eps,\delta'\rightarrow 0} \int_{C(\eps,\delta')}& \frac{\rmd f_n(1+\eps,\lambda,\alpha) }{\rmd \lambda} \frac{\Gamma(1+\beta(\lambda))^2}{\Gamma(-\beta(\lambda))^2} L_\star^{-4\beta(\lambda)}\rmd \lambda\nonumber\\
&=- 2\pi\rmi L_\star^{2-\frac{2}{n}(1-\alpha/\pi)}\frac{\Gamma(\frac{1}{2}+\frac{1}{2n}(1-\alpha/\pi))^2}{\Gamma(\frac{1}{2}-\frac{1}{2n}(1-\alpha/\pi))^2} (1+o(1)).
\end{align} 
Putting this all together, we have a correction term of the form given in \eqref{eq:subleading2}.

 \section{Vertex operators and Fisher-Hartwig representations for \texorpdfstring{$N=2$}{N=2}} \label{app:phases}
 In this appendix we look in more detail at corrections to the charged moment coming from the conjectured expansion of the Toeplitz determinant and the corresponding corrections coming from the CFT expansion. We will consider the case $N=2$. Then we have:
 \begin{align}
 \sigma = \Bigg(4\sin(k_1)\sin(k_2)\frac{\sin^2\left(\frac{k_2-k_1}{2}\right)}{\sin^2\left(\frac{k_2+k_1}{2}\right)}\Bigg)^{1/2}. \end{align}
 \subsection{Fisher-Hartwig terms}
 We can specialise \eqref{eq:subleading2} to the case $N=2$. Then:
 \begin{align}
d_n(\alpha,L) &= L^{-\frac{2}{n}(1+\alpha/\pi)}\frac{\Gamma(\frac{1}{2}+\frac{1}{2n}(1+\alpha/\pi))^2}{\Gamma(\frac{1}{2}-\frac{1}{2n}(1+\alpha/\pi))^2}\Bigg( c_1(-\alpha) \rme^{\rmi(k_2-k_1)L}  + c_2(-\alpha) \rme^{2\rmi k_2L}  + c_3(-\alpha) \rme^{-2\rmi k_1L} \Bigg)\nonumber \\
&+ L^{-\frac{2}{n}(1-\alpha/\pi)}\frac{\Gamma(\frac{1}{2}+\frac{1}{2n}(1-\alpha/\pi))^2}{\Gamma(\frac{1}{2}-\frac{1}{2n}(1-\alpha/\pi))^2}\Bigg( c_1(\alpha) \rme^{-\rmi(k_2-k_1)L}  + c_2(\alpha) \rme^{-2\rmi k_2L}  + c_3(\alpha) \rme^{2\rmi k_1L} \Bigg) +O(L^{-1})\label{eq:subleading3}\end{align} 
where:
\begin{align}
c_1(\alpha)&=2 \times \left(4\sin\left(\frac{k_2-k_1}{2}\right)\right)^{-\frac{2}{n}(1-\alpha/\pi)} \Bigg(\frac{\sin(k_1)\sin(k_2)}{\sin\left(\frac{k_2+k_1}{2}\right)}\Bigg)^{2-\frac{2}{n}(1-\alpha/\pi)}\nonumber\\
c_2(\alpha) &=(4\sin(k_2))^{-\frac{2}{n}(1-\alpha/\pi)} \Bigg(\frac{\sin\left(\frac{k_2-k_1}{2}\right)}{\sin\left(\frac{k_2+k_1}{2}\right)}\Bigg)^{2-\frac{2}{n}(1-\alpha/\pi)}\nonumber\\
c_3(\alpha) &=(4\sin(k_1))^{-\frac{2}{n}(1-\alpha/\pi)} \Bigg(\frac{\sin\left(\frac{k_2-k_1}{2}\right)}{\sin\left(\frac{k_2+k_1}{2}\right)}\Bigg)^{2-\frac{2}{n}(1-\alpha/\pi)}.
\end{align}
Note that the terms oscillating as $\rme^{\pm \rmi (k_2-k_1)L}$ each correspond to two summands in \eqref{eq:subleading2}. 
\subsection{CFT operators}
Now let us consider the CFT description for $N=2$. Then, considering the expansion \eqref{eq:descendants}, for $n\geq 2$ and $\alpha>0$ the dominant terms correspond to correlators of vertex operators: \begin{enumerate}
\item$ \rme^{\rmi (\frac{\alpha}{2\pi} -1) \varphi_1(x)}\rme^{\rmi \frac{\alpha}{2\pi} \varphi_2(x)}$,
\item  $\rme^{\rmi \frac{\alpha}{2\pi} \varphi_1(x)}\rme^{\rmi ( \frac{\alpha}{2\pi}-1) \varphi_2(x)}$,
\item  $\rme^{\rmi \left(\frac{\alpha}{2\pi}-\frac{1}{2}\right) \varphi_1(x)}\rme^{\rmi ( \frac{\alpha}{2\pi}-\frac{1}{2}) \varphi_2(x)}\rme^{\rmi \theta_1(x)}\rme^{ -\rmi\theta_2(x)}$,
\item $\rme^{\rmi \left(\frac{\alpha}{2\pi}-\frac{1}{2}\right) \varphi_1(x)}\rme^{\rmi ( \frac{\alpha}{2\pi}-\frac{1}{2}) \varphi_2(x)}\rme^{-\rmi \theta_1(x)}\rme^{\rmi\theta_2(x)}$.
\end{enumerate}
Using \eqref{eq:CFTformula}, the two-point function in the orbifold CFT of each of these operators scales as \begin{align}
 L^{-\frac{1}{3}(n-1/n)-\frac{4}{n}(\frac{\alpha}{2\pi})^2 -\frac{2}{n}\left(1-\frac{\alpha}{\pi}\right)},
 \end{align}
and so the correction to \eqref{eq:CFTmoment} is $o(1) = O(L^{-\frac{2}{n}\left(1-\frac{\alpha}{\pi}\right)})$. Now, the dominant term leading to \eqref{eq:CFTmoment} allows us to fix the phase factors that come with the fields $\varphi_k(x)$. In particular, we put $\varphi_1(x) \rightarrow \varphi_1(x) +2 k_1 x$ and $\varphi_2(x) \rightarrow \varphi_2(x) -2 k_2 x$. This means that in the two-point functions of the subdominant terms listed above, we have an oscillatory factor $\rme^{2\rmi k_1L}$ for (1), an oscillatory factor $ \rme^{-2\rmi k_2 L}$ for (2) and the correlator of both (3) and (4) oscillates as $\rme^{\rmi(k_1- k_2) L}$. An analogous discussion holds in the case $\alpha<0$.
\subsection{Conclusion}
We see that calculating the charged moment using an expansion of the Toeplitz determinant based on representations of the symbol with $\lvert\lvert\beta \rvert\rvert>1$, and calculating the charged moment using an expansion in CFT fields leads to the same order of subleading term. Moreover, the oscillatory terms that arise in the two methods coincide exactly. This correspondence is perhaps even closer: there are two vertex operators and two Fisher-Hartwig representations corresponding to $\rme^{\rmi(k_1- k_2) L}$, while there is one vertex operator and one Fisher-Hartwig representation corresponding to each of the other oscillatory terms. We point out that the lattice calculation \eqref{eq:subleading3} does not lead to a sum over terms of the form $L_\sigma^{-\frac{2}{n}\left(1\pm\frac{\alpha}{\pi}\right)}$, for the length-scale $\sigma$ given above. This means that in the CFT expansion \eqref{eq:descendants} the prefactors depend non-trivially on the Fermi momenta. This is in contrast to the $N=1$ XX model case, where the dependence is on $L_\sigma$ also in subleading terms.
\section{Leading descendant correction to charged moment}\label{app:descendant}
In this appendix we prove certain claims made in Section \ref{sec:descendantcorrections} about the correction of order $L^{-1}$ to the charged moment. 
First, recall the definition \eqref{eq:descendantintegral} of $J(n,\alpha)$:
\begin{align}
J(n,\alpha) = n \int_{-\infty}^\infty \left(\tanh(\pi w)-\tanh(n \pi w+\rmi\alpha/2) \right) (3w^2-1/4) \rmd w.\end{align}
We now use a trick that was used to evaluate \eqref{eq:a1} in \cite{Bonsignori19}. Writing $z= n \pi w$, for $\lvert \alpha \rvert <\pi$ we have:
\begin{align}
\frac{\rmd} {\rmd \alpha} J(n,\alpha) =-\frac{\rmi}{2\pi} \int_{-\infty}^\infty \frac{1}{\cosh(z + \rmi \alpha/2)^2} \Big(\frac{3}{n^2\pi^2}z^2-1/4\Big) \rmd z.\end{align}
The singularities of the integrand nearest to the real line are at $z +\rmi\alpha/2 = \pm \rmi \pi/2$, we can thus deform the contour to get:
\begin{align}
\frac{\rmd} {\rmd \alpha} J(n,\alpha) =-\frac{\rmi}{2\pi} \int_{-\infty}^\infty \frac{1}{\cosh(z)^2}  \Big(\frac{3}{n^2\pi^2}(z-\rmi\alpha/2)^2-1/4\Big) \ \rmd z=\frac{3\rmi }{4n^2\pi^3}\alpha^2 + \rmi\frac{1}{4\pi n^2}(n^2-1).\end{align}
Since $J(n,0) =0$, we can integrate to reach the claimed formula \eqref{eq:J}.

Now let us consider the correction of order $L_\sigma^{-1}$ for the full-counting statistics generating function, $Z_1(\alpha)$. We take the case $n=1$ in \eqref{eq:descendantintegral} to give:
\begin{align}
Z_1(\alpha)= \rme^{\rmi \alpha \langle Q_A\rangle + \Upsilon(1,\alpha)} L_\sigma^{-2\left(\frac{\alpha}{2\pi}\right)^2} \left(1- \frac{2\cos(k_F)}{ L_\sigma} J(1,\alpha)+o(L^{-1}) \right).\end{align}
Using \eqref{eq:J}, we have:
\begin{align}
Z_1(\alpha)= \rme^{\rmi \alpha \langle Q_A\rangle + \Upsilon(1,\alpha)} L_\sigma^{-2\left(\frac{\alpha}{2\pi}\right)^2} \left(1- \rmi\frac{\alpha^3}{2\pi^3}\frac{\cos(k_F)}{L_\sigma} +o(L^{-1}) \right).\end{align}
This result was conjectured in \cite{Abanov11} based on numerical calculations, and was then derived analytically, assuming the Fisher-Hartwig expansion \eqref{eq:fullexpansion}, in Ref.~\cite{Ivanov13b}. Since we use formula \eqref{eq:descendantintegral}, our derivation also implicitly assumes that this expansion holds. Alternatively, we could assume that the expansion given in \eqref{eq:subleading} is uniform.

\end{document}